# The Alan Turing Institute

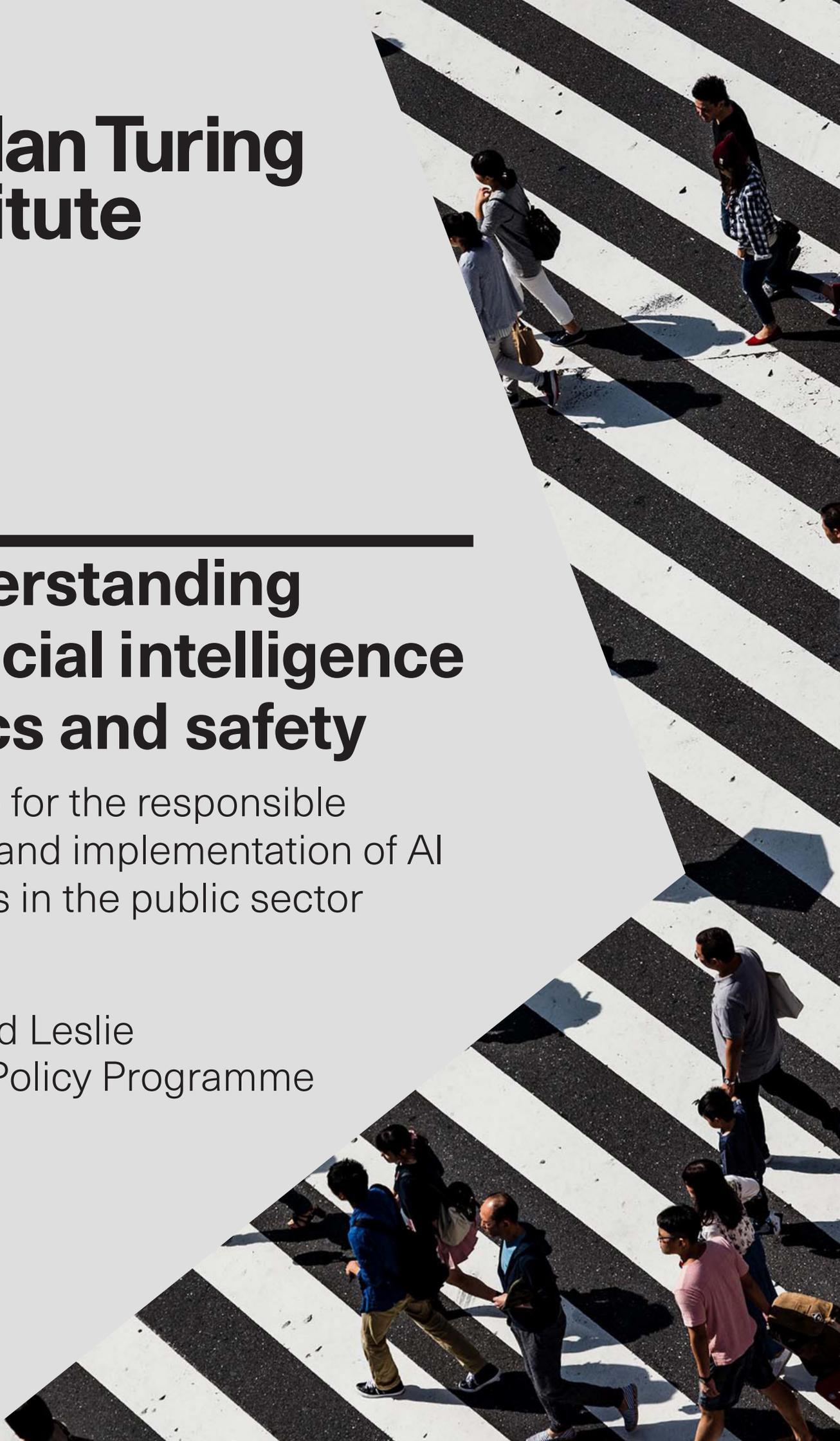

# Understanding artificial intelligence ethics and safety

A guide for the responsible design and implementation of AI systems in the public sector

Dr David Leslie
Public Policy Programme

**The
Alan Turing
Institute**

The Public Policy Programme at The Alan Turing Institute was set up in May 2018 with the
aim of developing research, tools, and techniques that help governments innovate with data-intensive
technologies and improve the quality of people's lives. We work alongside policy makers to explore how data
science and artificial intelligence can inform public policy and improve the provision of public services. We
believe that governments can reap the benefits of these technologies only if they make considerations of
ethics and safety a first priority.

This document provides end-to-end guidance on how to apply principles of AI ethics and safety to the design
and implementation of algorithmic systems in the public sector. We will shortly release a workbook to bring
the recommendations made in this guide to life. The workbook will contain case studies highlighting how the
guidance contained here can be applied to concrete AI projects. It will also contain exercises and practical
tools to help strengthen the process-based governance of your AI project.

Please note, that this guide is a living document that will evolve and improve with input from users, affected
stakeholders, and interested parties. We need your participation. Please share feedback with us at
policy@turing.ac.uk

This work was supported exclusively by the Turing's Public Policy Programme. All research undertaken by the
Turing's Public Policy Programme is supported entirely by public funds.

https://www.turing.ac.uk/research/research-programmes/public-policy

Cite this work as:

Leslie, D. (2019). Understanding artificial intelligence ethics and
safety: A guide for the responsible design and implementation of AI
systems in the public sector. *The Alan Turing Institute*.
https://doi.org/10.5281/zenodo.3240529

# Table of Contents:

## What is AI ethics?

### Intended audience and existing government guidance

The following guidance is designed to outline values, principles, and guidelines to assist department and delivery leads in ensuring that they develop and deploy AI ethically, safely, and responsibly. It is designed to complement and supplement the Data Ethics Framework. The Data Ethics Framework is a practical tool that should be used in any project initiation phase.

### AI ethics

A remarkable time of human promise has been ushered in by the convergence of the ever-expanding availability of big data, the soaring speed and stretch of cloud computing platforms, and the advancement of increasingly sophisticated machine learning algorithms.

This brave new digitally interconnected world is delivering rapid gains in the power of AI to better society. Innovations in AI are already dramatically improving the provision of essential social goods and services from healthcare, education, and transportation to food supply, energy, and environmental management. These bounties are, in fact, likely just the start. Because AI and machine learning systems organically improve with the enlargement of access to data and the growth of computing power, they will only become more effective and useful as the information age continues to develop apace. It may not be long before AI technologies become gatekeepers for the advancement of vital public interests and sustainable human development.

This prospect that progress in AI will help humanity to confront some of its most urgent challenges is exciting, but legitimate worries still abound. As with any new and rapidly evolving technology, a steep learning curve means that mistakes and miscalculations will be made and that both unanticipated and harmful impacts will inevitably occur.  AI is no exception.

In order to manage these impacts responsibly and to direct the development of AI systems toward optimal public benefit, you will have to make considerations of **AI ethics and safety a first priority**.

This will involve integrating considerations of the social and ethical implications of the design and use of AI systems into **every stage** of the delivery of your AI project. It will also involve a **collaborative effort** between the data scientists, product managers, data engineers, domain experts, and delivery managers on your team to align the development of artificial intelligence technologies with ethical values and principles that safeguard and promote the wellbeing of the communities that these technologies affect.

By including a primer on AI ethics with the Guide, we are providing you with the conceptual resources and practical tools that will enable you to steward the responsible design and implementation of AI projects.

 *AI ethics is a set of values, principles, and techniques that employ widely accepted standards of right and wrong to guide moral conduct in the development and use of AI technologies.*

These values, principles, and techniques are intended both to motivate morally acceptable practices and to prescribe the basic duties and obligations necessary to produce ethical, fair, and safe AI applications.

## Why AI ethics?

The field of AI ethics has largely emerged as a response to the range of individual and societal harms that the misuse, abuse, poor design, or negative unintended consequences of AI systems may cause. As a way to orient you to the importance of building a robust culture of AI ethics, here is a table that represents some of the most consequential forms that these potential harms may take:

| Potential Harms Caused by AI Systems |
| --- |
| **Bias and Discrimination** |
| Because they gain their insights from the existing structures and dynamics of the societies they analyse, data-driven technologies can reproduce, reinforce, and amplify the patterns of marginalisation, inequality, and discrimination that exist in these societies. |
| Likewise, because many of the features, metrics, and analytic structures of the models that enable data mining are chosen by their designers, these technologies can potentially replicate their designers' preconceptions and biases. |
| Finally, the data samples used to train and test algorithmic systems can often be insufficiently representative of the populations from which they are drawing inferences. This creates real possibilities of biased and discriminatory outcomes, because the data being fed into the systems is flawed from the start. |
| **Denial of Individual Autonomy, Recourse, and Rights** |
| When citizens are subject to decisions, predictions, or classifications produced by AI systems, situations may arise where such individuals are unable to hold directly accountable the parties responsible for these outcomes. |
| AI systems automate cognitive functions that were previously attributable exclusively to accountable human agents. This can complicate the designation of responsibility in algorithmically generated outcomes, because the complex and distributed character of the design, production, and implementation processes of AI systems may make it difficult to pinpoint accountable parties. |
| In cases of injury or negative consequence, such an accountability gap may harm the autonomy and violate the rights of the affected individuals. |
| **Non-transparent, Unexplainable, or Unjustifiable Outcomes** |
| Many machine learning models generate their results by operating on high dimensional correlations that are beyond the interpretive capabilities of human scale reasoning. In these cases, the rationale of algorithmically produced outcomes that directly affect decision subjects remains opaque to those subjects. While in some use cases, this lack of explainability may be acceptable, in some applications, where the processed data could |

harbour traces of discrimination, bias, inequity, or unfairness, the opaqueness of the model may be deeply problematic.

## Invasions of Privacy

Threats to privacy are posed by AI systems both as a result of their design and development processes, and as a result of their deployment. As AI projects are anchored in the structuring and processing of data, the development of AI technologies will frequently involve the utilisation of personal data. This data is sometimes captured and extracted without gaining the proper consent of the data subject or is handled in a way that reveals (or places under risk the revelation of) personal information.

On the deployment end, AI systems that target, profile, or nudge data subjects without their knowledge or consent could in some circumstances be interpreted as infringing upon their ability to lead a private life in which they are able to intentionally manage the transformative effects of the technologies that influence and shape their development. This sort of privacy invasion can consequently harm a person's more basic right to pursue their goals and life plans free from unchosen influence.

## Isolation and Disintegration of Social Connection

While the capacity of AI systems to curate individual experiences and to personalise digital services holds the promise of vastly improving consumer life and service delivery, this benefit also comes with potential risks. Excessive automation, for example, might reduce the need for human-to-human interaction, while algorithmically enabled hyper-personalisation, by limiting our exposure to worldviews different from ours, might polarise social relationships. Well-ordered and cohesive societies are built on relations of trust, empathy, and mutual understanding. As AI technologies become more prevalent, it is important that these relations be preserved.

## Unreliable, Unsafe, or Poor-Quality Outcomes

Irresponsible data management, negligent design and production processes, and questionable deployment practices can, each in their own ways, lead to the implementation and distribution of AI systems that produce unreliable, unsafe, or poor-quality outcomes. These outcomes can do direct damage to the wellbeing of individual persons and the public welfare. They can also undermine public trust in the responsible use of societally beneficial AI technologies, and they can create harmful inefficiencies by virtue of the dedication of limited public resources to inefficient or even detrimental AI technologies.

## An ethical platform for the responsible delivery of an AI project

Building a project delivery environment, which enables the ethical design and deployment of AI systems, requires a multidisciplinary team effort. It demands the active cooperation of all team members both in maintaining a **deeply ingrained culture of responsibility** and in executing a **governance architecture that adopts ethically sound practices at every point in the innovation and implementation lifecycle**.

This task of uniting an in-built culture of responsible innovation with a governance architecture that brings the values and principles of ethical, fair, and safe AI to life, will require that you and your team accomplish several goals:

- You will have to ensure that your AI project is *ethically permissible* by considering the impacts it may have on the wellbeing of affected stakeholders and communities.

- You will have to ensure that your AI project is **fair and non-discriminatory** by accounting for its potential to have discriminatory effects on individuals and social groups, by mitigating biases that may influence your model's outputs, and by being aware of the issues surrounding fairness that come into play at every phase of the design and implementation pipeline.

- You will have to ensure that your AI project is **worthy of public trust** by guaranteeing to the extent possible the safety, accuracy, reliability, security, and robustness of its product.

- You will have to ensure that your AI project is **justifiable** by prioritising both the transparency of the process by which your model is designed and implemented, and the transparency and interpretability of its decisions and behaviours.

We call this governance architecture an ***ethical platform*** for two important reasons. First, it is intended to provide you with a solid, process-based footing of values, principles, and protocols—*an ethical platform to stand on*—so that you and your team are better able to design and implement AI systems ethically, equitably, and safely. Secondly, it is intended to help you facilitate a culture of responsible AI innovation—*to help you provide an ethical platform to stand for*—so that your project team can be united in a collaborative spirit to develop AI technologies for the public good.

## Preliminary considerations about the ethical platform

Our aim for the remainder of this document is to provide you with guidance that is as comprehensive as possible in its presentation of the values, principles, and governance mechanisms necessary to serve the purpose of responsible innovation. Keep in mind, however, that not all issues discussed in this document will apply equally to each project. Clearly, a machine learning algorithm trained to detect spam emails will present fewer ethical challenges compared to one trained to detect cancer in blood samples. Similarly, image recognition systems used for sorting and routing mail raise fewer ethical dilemmas compared to the facial recognition technologies used in law enforcement.

Low-stakes AI applications that are not safety-critical, do not directly impact the lives of people, and do not process potentially sensitive social and demographic data will need less proactive ethical stewardship than high-stakes projects. You and your project team will need to evaluate the scope and possible impacts of your project on affected individuals and communities, and you will have to apply reasonable assessments of the risks posed to individual wellbeing and public welfare in order to formulate proportional governance procedures and protocols.

Be that as it may, you should also keep in mind that all AI projects have social and ethical impacts on stakeholders and communities even if just by diverting or redistributing limited intellectual, material, and economic resources away from other concerns and possibilities for socially beneficial innovation. Ethical considerations and principles-based policy formation should therefore play a salient role in every prospective AI project.

## Three building-blocks of a responsible AI project delivery ecosystem

Setting up an ethical platform for responsible AI project delivery involves not only **building from the cultural ground up**; it involves providing your team with the means to accomplish the goals of establishing the ethical permissibility, fairness, trustworthiness, and justifiability of your project. It will take three building-blocks to make such an ethical platform possible:

1. At the most basic level, it necessitates that you gain a working knowledge of a framework of **ethical values that *Support, Underwrite, and Motivate*** a responsible data design and use ecosystem. These will be called **SUM Values**, and they will be composed of four key notions: ***Respect, Connect, Care, and Protect***. The objectives of these SUM Values are (1) to provide you with an accessible framework to start thinking about the moral scope of the societal and ethical impacts of your project and (2) to establish well-defined criteria to evaluate its ethical permissibility.

2. At a second and more concrete level, an ethical platform for responsible AI project delivery requires a set of **actionable principles** that facilitate an orientation to the responsible design and use of AI systems. These will be called **FAST Track Principles**, and they will be composed of four key notions: ***Fairness, Accountability, Sustainability, and Transparency***. The objectives of these FAST Track Principles are to provide you with the moral and practical tools (1) to make sure that your project is bias-mitigating, non-discriminatory, and fair, and (2) to safeguard public trust in your project's capacity to deliver safe and reliable AI innovation.

3. At a third and most concrete level, an ethical platform for responsible AI project delivery requires a **process-based governance framework (PBG Framework)** that **operationalises the SUM Values and the FAST Track Principles** across the entire AI project delivery workflow. The objective of this PBG Framework is to set up transparent processes of design and implementation that safeguard and enable the justifiability of both your AI project and its product.

Here is a summary visualisation of these three building blocks of the platform:

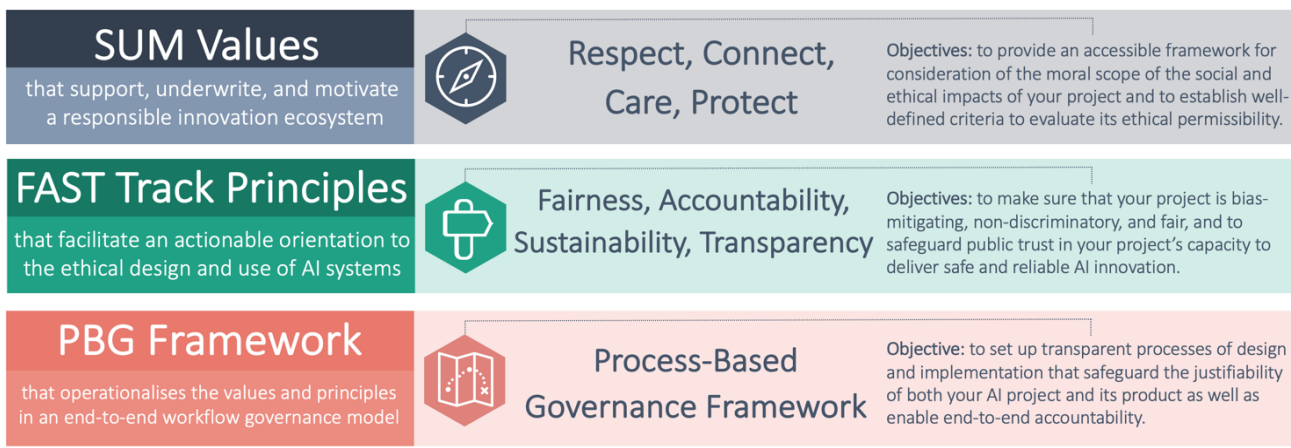

## Ethical Platform for the Responsible Delivery of an AI Project

**SUM Values**
that support, underwrite, and motivate a responsible innovation ecosystem

Respect, Connect, Care, Protect

Objectives: to provide an accessible framework for consideration of the moral scope of the social and ethical impacts of your project and to establish well-defined criteria to evaluate its ethical permissibility.

**FAST Track Principles**
that facilitate an actionable orientation to the ethical design and use of AI systems

Fairness, Accountability, Sustainability, Transparency

Objectives: to make sure that your project is bias-mitigating, non-discriminatory, and fair, and to safeguard public trust in your project's capacity to deliver safe and reliable AI innovation.

**PBG Framework**
that operationalises the values and principles in an end-to-end workflow governance model

Process-Based Governance Framework

Objective: to set up transparent processes of design and implementation that safeguard the justifiability of both your AI project and its product as well as enable end-to-end accountability.

*How to use this guide*

This guide is intended to assist you in stewarding practices of responsible AI innovation. This entails that the ethical platform be put into practice at every step of the design and implementation workflow. Turning the SUM Values, the FAST Track Principles, and the PBG Framework into practice will require that you and your team continuously **reflect, act, and justify**:

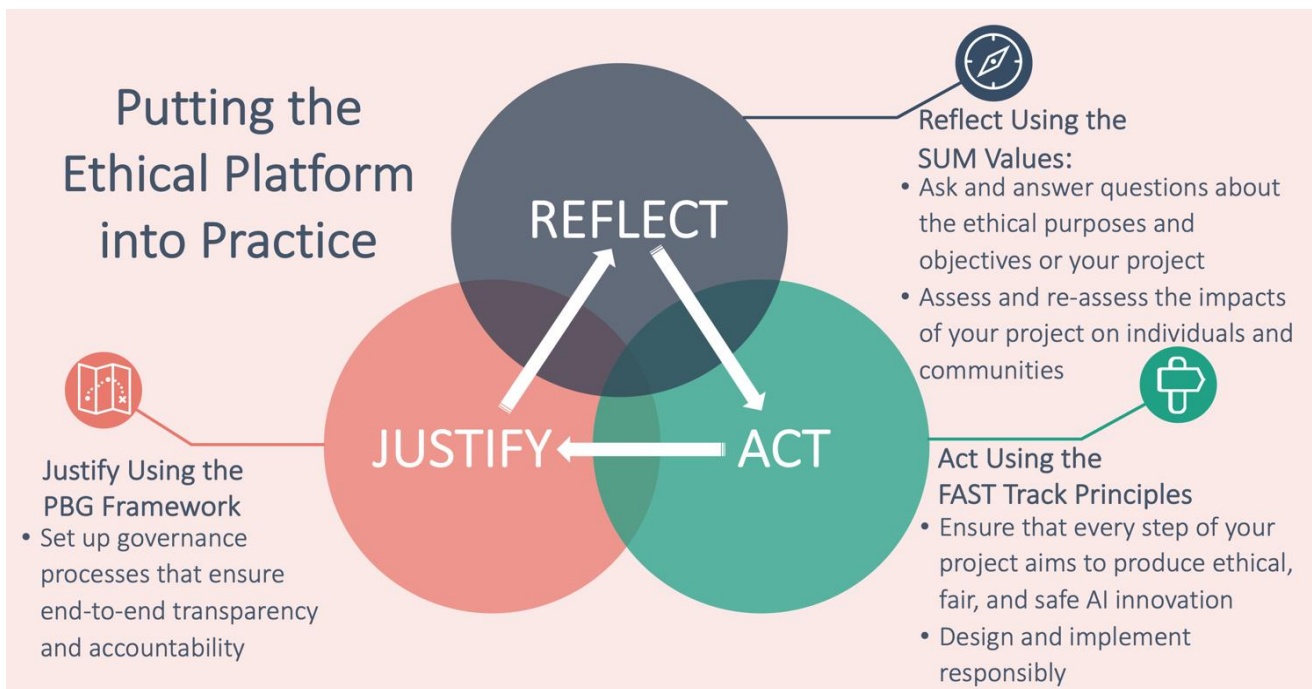

## The SUM Values

### Background

The challenge of creating a culture of responsible innovation begins with the task of building an **accessible moral vocabulary** that will allow team members to explore and discuss the ethical stakes of the AI projects that they are involved in or are considering taking on.

In the field of AI ethics, this moral vocabulary draws primarily on two traditions of moral thinking: (1) **bioethics** and (2) **human rights discourse. Bioethics** is the study of the ethical impacts of biomedicine and the applied life sciences. **Human rights discourse** draws inspiration from the UN Declaration of Human Rights. It is anchored in a set of universal principles that build upon the idea that all humans have an equal moral status as bearers of intrinsic human dignity.

Whereas bioethics largely stresses the normative values that underlie the safeguarding of *individuals* in instances where technological practices affect their interests and wellbeing, human rights discourse mainly focuses on the set of **social, political, and legal entitlements** that are due to all human beings under a universal framework of juridical protection and the rule of law.

The main principles of bioethics include **respecting the autonomy of the individual, protecting people from harm, looking after the well-being of others, and treating all individuals equitably and justly**. The main tenets of human rights include **the entitlement to equal freedom and dignity under the law, the protection of civil, political, and social rights, the universal recognition of personhood, and the right to free and unencumbered participation in the life of the community**.

*The SUM Values: Respect, Connect, Care, and Protect*

While the SUM Values incorporate conceptual elements from both bioethics and human rights discourse, they do so with an eye to applying the most critical of these elements to the specific social and ethical problems raised by the potential misuse, abuse, poor design, or harmful unintended consequences of AI systems.

They are also meant to be utilised as guiding values throughout the innovation lifecycle: from the preliminary steps of project evaluation, planning, and problem formulation, through processes of design, development, and testing, to the stages of implementation and reassessment. The SUM Values can be visualised as follows:

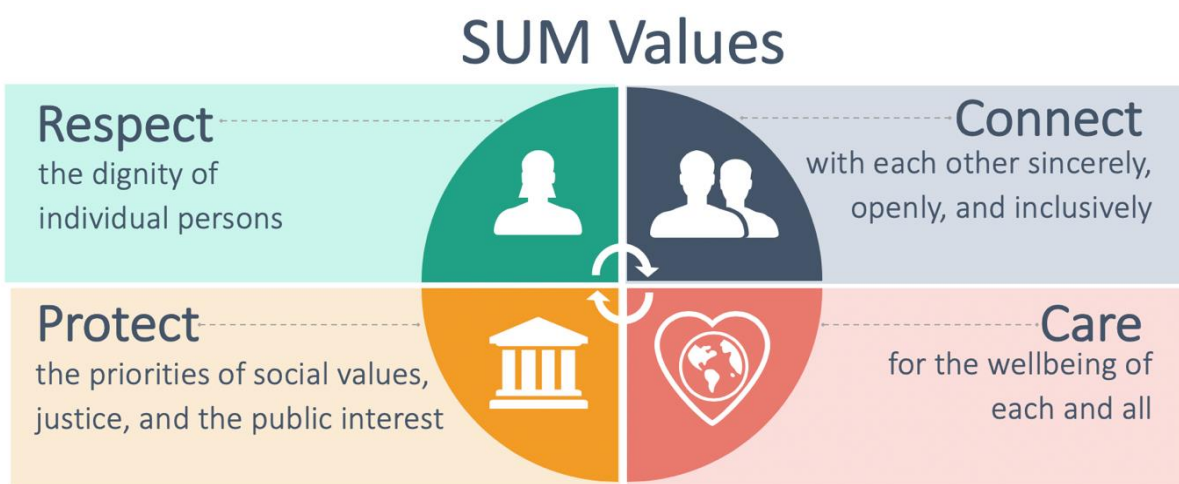

In order to focus in on a more detailed exploration of each of the values' meanings, their contents will be presented individually. Formulating it as a question: What are each of these values charging you to do?

→ RESPECT the dignity of individual persons:

- Ensure their abilities to make free and informed decisions about their own lives

- Safeguard their autonomy, their power to express themselves, and their right to be heard

- Secure their capacities to make well-considered and independent contributions to the life of the community

- Support their abilities to flourish, to fully develop themselves, and to pursue their passions and talents according to their own freely determined life plans

→ CONNECT with each other sincerely, openly, and inclusively:

- Safeguard the integrity of interpersonal dialogue, meaningful human connection, and social cohesion

- Prioritise diversity, participation, and inclusion at all points in the design, development, and deployment processes of AI innovation.

- Encourage all voices to be heard and all opinions to be weighed seriously and sincerely throughout the production and use lifecycle

- Use the advancement and proliferation of AI technologies to strengthen the developmentally essential relationship between interacting human beings.

- Utilise AI innovations *pro-socially* so as to enable bonds of interpersonal solidarity to form and individuals to be socialised and recognised by each other

- Use AI technologies to foster this capacity to connect so as to reinforce the edifice of trust, empathy, reciprocal responsibility, and mutual understanding upon which all ethically well-founded social orders rest

→ CARE for the wellbeing of each and all:

- Design and deploy AI systems to foster and to cultivate the welfare of all stakeholders whose interests are affected by their use

- Do no harm with these technologies and minimise the risks of their misuse or abuse

- Prioritise the safety and the mental and physical integrity of people when scanning horizons of technological possibility and when conceiving of and deploying AI applications

→ PROTECT the priorities of social values, justice, and the public interest:

- Treat all individuals equally and protect social equity

- Use digital technologies as an essential support for the protection of fair and equal treatment under the law

- Prioritise social welfare, public interest, and the consideration of the social and ethical impacts of innovation in determining the legitimacy and desirability of AI technologies

- Use AI to empower and to advance the interests and well-being of as many individuals as possible

- Think big-picture about the wider impacts of the AI technologies you are conceiving and developing. Think about the ramifications of their effects and externalities for others around the globe, for future generations, and for the biosphere as a whole

As a general rule, these SUM Values should orient you in deliberating about the **ethical permissibility** of a prospective AI project. They should also provide you with a framework of concepts to consider the **ethical impacts of an AI system across the design, use, and monitoring lifecycle**.

Taking these SUM Values as a starting point of conversation, you should also encourage discussion within your team of how to weigh the values against one another and how to consider trade-offs should use case specific circumstances arise when the values come into tension with each other.

## The FAST Track Principles:

### Background

While the SUM Values are intended to provide you with some general normative guideposts and moral motivations for thinking through the social and ethical aspects of AI project delivery, they are not specifically catered to the actual processes involved in developing and deploying AI systems.

To make clear what is needed for this next step toward a more actionable orientation to the responsible design and use of AI technologies, it would be helpful to briefly touch upon what has necessitated the emergence of AI ethics in the first place.

Marvin Minsky, the great cognitive scientist and AI pioneer, defined AI as follows: 'Artificial Intelligence is the science of *making computers do things that require intelligence* when done by humans.' This standard definition should key us in to the principal motivation that has driven the development of the field of the applied ethics of artificial intelligence:

When humans do 'things that require intelligence,' we hold them responsible for the accuracy, reliability, and soundness of their judgements. Moreover, we demand of them that their actions and decisions be supported by good reasons, and we hold them accountable for the fairness, equity, and reasonableness of how they treat others.

What creates the need for principles tailored to the design and use of AI systems is that their emergence and expanding power 'to do things that require intelligence' has heralded a shift of a wide array of cognitive functions to algorithmic processes that themselves can be held neither directly responsible nor immediately accountable for the consequences of their behaviour.

As inert and program-based machinery, AI systems are not morally accountable agents. This has created an ethical breach in the sphere of the applied science of AI that the growing number of frameworks for AI ethics are currently trying to fill. Targeted principles such as fairness, accountability, sustainability, and transparency are meant to 'fill the gap' between the new 'smart agency' of machines and their fundamental lack of moral responsibility.

### The FAST Track Principles: Fairness, Accountability, Sustainability, and Transparency

By becoming well-acquainted with the FAST Track Principles, *all members* of your project delivery team will be better able to support a responsible environment for data innovation.

Issues of fairness, accountability, sustainability, and transparency operate at every juncture and at every level of the AI project delivery workflow and demand the cooperative attention and deliberative involvement of those with technical expertise, domain knowledge, project/product management skill, and policy competence. Ethical AI innovation is a team effort from start to finish.

To introduce you to the scope of the FAST Track Principles, here is a summary visualisation of them:

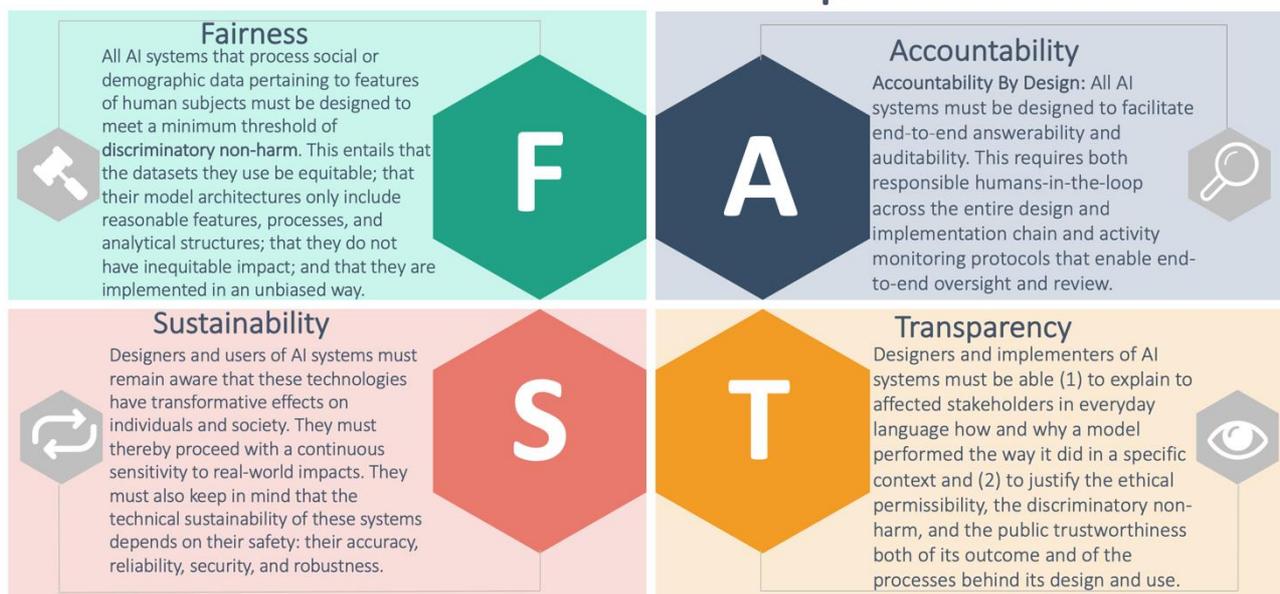

You should keep in mind, initially, that while fairness, accountability, sustainability, and transparency are grouped together in the FAST acronym, they do not necessarily relate to each other on the same plane or as equivalents. The principles of accountability and transparency are ***end-to-end governing principles***. Accountability entails that humans are answerable for the parts they play across the entire AI design and implementation workflow. It also demands that the results of this work are traceable from start to finish. The principle of transparency entails that design and implementation processes are justifiable through and through. It demands as well that an algorithmically influenced outcome is interpretable and made understandable to affected parties.

The governing roles of accountability and transparency are very different from the more dependent roles of fairness and sustainability. These latter two are *qualities* of algorithmic systems for which their designers and implementers are ***held accountable*** through the ***transparency both of the outcomes of their practices and of the practices themselves***. According to the principle of fairness, designers and implementers are held accountable for being equitable and for not harming anyone through bias or discrimination.  According to the principle of sustainability, designers and implementers are held accountable for producing AI innovation that is safe and ethical in its outcomes and wider impacts.

Whereas the principles of transparency and accountability thus provide the procedural mechanisms and means through which AI systems can be justified and by which their producer and implementers can be held responsible, fairness and sustainability are the crucial aspects of the design, implementation, and outcomes of these systems which establish the normative criteria for such governing constraints.  These four principles are therefore all deeply interrelated, but they are not equal.

There is one more important thing to keep in mind before we delve into the details of the FAST Track principles. Transparency, accountability, and fairness are *also* ***data protection principles***, and where algorithmic processing involves personal data, complying with them is not simply a matter of ethics or good practice, but a legal requirement, which is enshrined in the General Data Protection Regulation (GDPR) and the Data Protection Act of 2018 (DPA 2018). For more detailed information about the specific meanings of transparency, accountability, and fairness as data protection principles in the context of the GDPR and the DPA 2018, please refer to the  Guide to Data Protection produced by the Information Commissioner's Office.

## Fairness

When thinking about fairness in the design and deployment of AI systems, it is important to always keep in mind that these technologies, no matter how neutral they may seem, are designed and produced by human beings, who are bound by the limitations of their contexts and biases.

Human error, prejudice, and misjudgement can enter into the innovation lifecycle and create biases at any point in the project delivery process from the preliminary stages of data extraction, collection, and pre-processing to the critical phases of problem formulation, model building, and implementation.

Additionally, data-driven technologies achieve accuracy and efficacy by building inferences from datasets that record complex social and historical patterns, which themselves may contain culturally crystallised forms of bias and discrimination. There is no silver bullet when it comes to remediating the dangers of discrimination and unfairness in AI systems. The problem of fairness and bias mitigation in algorithmic design and use therefore has no simple or strictly technical solution.

That said, best practices of fairness-aware design and implementation (both at the level of non-technical self-assessment and at the level of technical controls and means of evaluation) hold great promise in terms of securing just, morally acceptable, and beneficial outcomes that treat affected stakeholders fairly and equitably.

While there are different ways to characterise or define fairness in the design and use of AI systems, you should consider the **principle of discriminatory non-harm** as a minimum required threshold of fairness. This principle directs us to do no harm to others through the biased or discriminatory outcomes that may result from practices of AI innovation:

---

**Principle of Discriminatory Non-Harm:** The designers and users of AI systems, which process social or demographic data pertaining to features of human subjects, societal patterns, or cultural formations, should prioritise the mitigation of bias and the exclusion of discriminatory influences on the outputs and implementations of their models. Prioritising discriminatory non-harm implies that the designers and users of AI systems ensure that the decisions and behaviours of their models do not generate discriminatory or inequitable impacts on affected individuals and communities. This entails that these designers and users ensure that the AI systems they are developing and deploying:

1. Are trained and tested on properly representative, relevant, accurate, and generalisable datasets (**Data Fairness**)
2. Have model architectures that do not include target variables, features, processes, or analytical structures (correlations, interactions, and inferences) which are unreasonable, morally objectionable, or unjustifiable (**Design Fairness**)
3. Do not have discriminatory or inequitable impacts on the lives of the people they affect (**Outcome Fairness**)
4. Are deployed by users sufficiently trained to implement them responsibly and without bias (**Implementation Fairness**)

---

*Data fairness*

Responsible data acquisition, handling, and management is a necessary component of algorithmic fairness. If the results of your AI project are generated by biased, compromised, or skewed datasets, affected stakeholders will not adequately be protected from discriminatory harm. Your project team should keep in mind the following key elements of data fairness:

- **Representativeness**: Depending on the context, either underrepresentation or overrepresentation of disadvantaged or legally protected groups in the data sample may lead to the systematic disadvantaging of vulnerable stakeholders in the outcomes of the trained model. To avoid such kinds of sampling bias, domain expertise will be crucial to assess the fit between the data collected or procured and the underlying population to be modelled. Technical team members should, if possible, offer means of remediation to correct for representational flaws in the sampling.

- **Fit-for-Purpose and Sufficiency**: An important question to consider in the data collection and procurement process is: Will the amount of data collected be sufficient for the intended purpose of the project? The quantity of data collected or procured has a significant impact on the accuracy and reasonableness of the outputs of a trained model. A data sample not large enough to represent with sufficient richness the significant or qualifying attributes of the members of a population to be classified may lead to unfair outcomes. Insufficient datasets may not equitably reflect the qualities that should rationally be weighed in producing a justified outcome that is consistent with the desired purpose of the AI system. Members of the project team with technical and policy competences should collaborate to determine if the data quantity is, in this respect, sufficient and fit-for-purpose.

- **Source Integrity and Measurement Accuracy:** Effective bias mitigation begins at the very commencement of data extraction and collection processes. Both the sources and instruments of measurement may introduce discriminatory factors into a dataset. When incorporated as inputs in the training data, biased prior human decisions and judgments—such as prejudiced scoring, ranking, interview-data or evaluation—will become the 'ground truth' of the model and replicate the bias in the outputs of the system. In order to secure discriminatory non-harm, you must do your best to make sure your data sample has optimal source integrity. This involves securing or confirming that the data gathering processes involved suitable, reliable, and impartial sources of measurement and sound methods of collection.

- **Timeliness and Recency:** If your datasets include outdated data then changes in the underlying data distribution may adversely affect the generalisability of your trained model. Provided these distributional drifts reflect changing social relationship or group dynamics, this loss of accuracy with regard to the actual characteristics of the underlying population may introduce bias into your AI system. In preventing discriminatory outcomes, you should scrutinise the timeliness and recency of all elements of the data that constitute your datasets.

- **Relevance, Appropriateness and Domain Knowledge**:  The understanding and utilisation of the most appropriate sources and types of data are crucial for building a robust and unbiased AI system. Solid domain knowledge of the underlying population distribution and of the predictive or classificatory goal of the project is instrumental for choosing optimally relevant measurement inputs that contribute to the reasonable determination of the defined solution. You should make sure that domain experts collaborate closely with your technical team to assist in the determination of the optimally appropriate categories and sources of measurement.

To ensure the uptake of best practices for responsible data acquisition, handling, and management across your AI project delivery workflow, you should initiate the creation of a **Dataset Factsheet** at the alpha stage of your project. This factsheet should be maintained diligently throughout the design and implementation lifecycle in order to secure optimal data quality, deliberate bias-mitigation aware practices, and optimal auditability. It should include **a comprehensive record of data provenance, procurement, pre-processing, lineage, storage, and security as well as qualitative input from team members about determinations made with regard to data representativeness, data sufficiency, source integrity, data timeliness, data relevance, training/testing/validating splits, and unforeseen data issues encountered across the workflow**.

*Design Fairness*

Because human beings have a hand in all stages of the construction of AI systems, fairness-aware design must take precautions across the AI project workflow to prevent bias from having a discriminatory influence:

- **Problem Formulation:** At the initial stage of problem formulation and outcome definition, technical and non-technical members of your team should work together to translate project goals into measurable targets. This will involve the use of both domain knowledge and technical understanding to define what is being optimised in a formalisable way and to translate the project's objective into a target variable or measurable proxy, which operates as a statistically actionable rendering of the defined outcome.

  At each of these points, choices must be made about the design of the algorithmic system that may introduce structural biases which ultimately lead to discriminatory harm. Special care must be taken here to identify affected stakeholders and to consider how vulnerable groups might be negatively impacted by the specification of outcome variables and proxies. Attention must also be paid to the question of whether these specifications are reasonable and justifiable given the general purpose of the project and the potential impacts that the outcomes of the system's use will have on the individuals and communities involved.

  These challenges of fairness aware design at the problem formulation stage show the need for making diversity and inclusive participation a priority from the start of the AI project lifecycle. This involves both the collaboration of the entire team and the attainment of stakeholder input about the acceptability of the project plan. This also entails collaborative deliberation across the project team and beyond about the ethical impacts of the design choices made.

- **Data Pre-Processing:** Human judgment enters into the process of algorithmic system construction at the stage of labelling, annotating, and organising the training data to be utilised in building the model. Choices made about how to classify and structure raw inputs must be taken in a fairness aware manner with due consideration given to the sensitive social contexts that may introduce bias into such acts of classification. Similar fairness aware processes should be put in place to review automated or outsourced classifications. Likewise, efforts should be made to attach solid contextual information and ample metadata to the datasets, so that downstream analyses of data processing have access to properties of concern in bias mitigation.

- **Feature Determination and Model-Building:** The constructive task of selecting the attributes or features that will serve as input variables for your model involves human decisions be made about what sorts of information may or may not be relevant or rationally required to yield an accurate *and* unbiased classification or prediction. Moreover, the feature engineering tasks of aggregating, extracting, or decomposing attributes from datasets may introduce human appraisals that have biasing effects. For this reason, discrimination awareness should play a large role at this stage of the AI model-building workflow as should domain knowledge and policy expertise. Your team should proceed in the modelling stage aware that choices made about grouping or separating and including or excluding features as well as more general judgements about the comprehensiveness or coarseness of the total set of features may have significant consequences for vulnerable or protected groups.

  The process of tuning hyperparameters and setting metrics at the modelling, testing, and evaluation stages also involves human choices that may have discriminatory effects in the trained model. Your technical team should proceed with an attentiveness to bias risk, and continual iterations of peer review and project team consultation should be encouraged to ensure that choices made in adjusting the dials and metrics of the model are in line with bias mitigation and discriminatory non-harm.

- **Evaluating Analytical Structures:** Design fairness also demands close assessment of the existence in the trained model of lurking or hidden proxies for discriminatory features that may act as significant factors in its output. Including such hidden proxies in the structure of the model may lead to implicit 'redlining' (the unfair treatment of a sensitive group on the basis of an unprotected attribute or interaction of attributes that 'stands in' for a protected or sensitive one).

  Designers must additionally scrutinise the moral justifiability of the significant correlations and inferences that are determined by the model's learning mechanisms themselves. In cases of the processing of social or demographic data related to human features, where the complexity and high dimensionality of machine learning models preclude the confirmation of the discriminatory non-harm of these inferences (for reason of their uninterpretability by human assessors), these models should be avoided. In AI systems that process and draw analytics from data arising from human relationships, societal patterns, and complex socioeconomic and cultural formations, designers must prioritise a degree of interpretability that is sufficient to ensure that the inferences produced by these systems are non-discriminatory. In cases where this is not possible, a different, more transparent and explainable model or portfolio of models should be chosen.

  Analytical structures must also be confirmed to be *procedurally fair*. Any rule or procedure employed in an AI system should be consistently and uniformly applied to every decision subject whose information is being processed by that system. Your team should be able to certify that when a rule or procedure has been used to render an outcome for any given individual, the same rule or procedure will be applied to any other individual in the same way regardless of that other subject's similarities with or differences from the first.

Implementers, in this respect, should be able to show that any algorithmic output is replicable when the same rules and procedures are applied to the same inputs. Such a uniformity of the application of rules and procedures secures the equal procedural treatment of decision subjects and precludes any rule-changes in the algorithmic processing targeted at a specific person that may disadvantage that individual vis-à-vis any other.

*Outcome fairness*

As part of this minimum safeguarding of discriminatory non-harm, forethought and well-informed consideration must be put into *how you are going to define and measure the fairness of the impacts and outcomes of the AI system you are developing*.

There is a great diversity of beliefs in the area of **outcome fairness** as to how to properly classify what makes the consequences of an algorithmically supported decision equitable, fair, and allocatively just. Different approaches—detailed below—stress different principles: some focus on demographic parity, some on individual fairness, others on error rates equitably distributed across subpopulations.

Your determination of outcome fairness should heavily depend both on the **specific use case for which the fairness of outcome is being considered** and the **technical feasibility of incorporating your chosen criteria into the construction of the AI system.** (Note that different fairness-aware methods involve different types of technical interventions at the pre-processing, modelling, or post-processing stages of production). Again, this means that determining your fairness definition should be a **cooperative and multidisciplinary effort across the project team**.

You will find below a summary table of some of the main definitions of outcome fairness that have been integrated by researchers into formal models as well as a list of current articles and technical resources, which should be consulted to orient your team to the relevant knowledge base. (Note that this is a rapidly developing field, so your technical team should keep updated about further advances.) The first four fairness types fall under the category of group fairness and allow for comparative criteria of non-discrimination to be considered in model construction and evaluation. The final two fairness types focus instead on cases of individual fairness, where context-specific issues of effective bias are considered and assessed at the level of the individual agent.

Take note, though, that these technical approaches have limited scope in terms of the bigger picture issues of algorithmic fairness that we have already stressed. Many of the formal approaches work only in use cases that have *distributive or allocative consequences*. In order to carry out group comparisons, these approaches require access to data about sensitive/protected attributes (that may often be unavailable or unreliable) as well as accurate demographic information about the underlying population distribution. Furthermore, there are unavoidable trade-offs and inconsistences between these technical definitions that must be weighed in determining which of them are best fit for your use case. Consult those on your project team with the technical expertise to consider the use case appropriateness of a desired formal approach.

| Some Formalisable Definitions of Outcome Fairness | |
|---|---|
| Type of Fairness | Definition |
| **Demographic/ Statistical Parity**<br><br>*Group Fairness* | An outcome is fair if each group in the selected set receives benefit in equal or similar proportions, i.e. if there is no correlation between a sensitive or protected attribute and the allocative result. This approach is intended to prevent *disparate impact*, which occurs when the outcome of an algorithmic process disproportionately harms members of disadvantaged or protected groups. |
| **True Positive Rate Parity**<br><br>*Group Fairness* | An outcome is fair if the 'true positive' rates of an algorithmic prediction or classification are equal across groups. This approach is intended to align the goals of bias mitigation and accuracy by ensuring that the accuracy of the model is equivalent between relevant population subgroups. This method is also referred to as 'equal opportunity' fairness because it aims to secure equalised odds of an advantageous outcome for qualified individuals in a given population regardless of the protected or disadvantaged groups of which they are members. |
| **False Positive Rate Parity**<br><br>*Group Fairness* | An outcome is fair if it does not disparately mistreat people belonging to a given social group by misclassifying them at a higher rate than the members of a second social group, for this would place the members of the first group at an unfair disadvantage. This approach is motivated by the position that sensitive groups and advantaged groups should have similar error rates in outcomes of algorithmic decisions. |
| **Positive Predictive Value Parity**<br><br>*Group Fairness* | An outcome is fair if the rates of positive predictive value (the fraction of correctly predicted positive cases out of all predicted positive cases) are equal across sensitive and advantaged groups. Outcome fairness is defined here in terms of a parity of precision, where the probability of members from different groups actually having the quality they are predicted to have is the same across groups. |
| **Individual Fairness**<br><br>*Individual Fairness* | An outcome is fair if it treats individuals with similar relevant qualifications similarly. This approach relies on the establishment of a similarity metric that shows the degree to which pairs of individuals are alike with regard to a specific task. |
| **Counterfactual Fairness**<br><br>*Individual Fairness* | An outcome is fair if an automated decision made about an individual belonging to a sensitive group would have been the same were that individual a member of a different group in a closest possible alternative (or counterfactual) world. Like the individual fairness approach, this method of defining fairness focuses on the specific circumstances of an affected decision subject, but, by using the tools of contrastive explanation, it moves beyond individual fairness insofar as it brings out the causal influences behind the algorithmic output. It also presents the possibility of offering the subject of an automated decision knowledge of what factors, if changed, could have influenced a different outcome. This could provide them with actionable recourse to change an unfavourable decision. |

Fairness Position Statement:

Once you and your project team have thoroughly considered the use case appropriateness as well as technical feasibility of the formal models of fairness most relevant for your system and have incorporated the model into your application, you should prepare a **Fairness Position Statement (FPS)** in which the fairness criteria being employed in the AI system is made explicit and explained in plain and non-technical language. This FPS should then be made publicly available for review by all affected stakeholders.

*Implementation fairness*

When your project team is approaching the beta stage, you should begin to build out your plan for implementation training and support. This plan should include adequate preparation for the responsible and unbiased deployment of the AI system by its on-the-ground users. Automated decision-support systems present novel risks of bias and misapplication at the point of delivery, so special attention should be paid to preventing harmful or discriminatory outcomes at this critical juncture of the AI project lifecycle.

In order to design an optimal regime of implementer training and support, you should pay special attention to the unique pitfalls of bias-in-use to which the deployment of AI technologies give rise. These can be loosely classified as decision-automation bias (more commonly just 'automation bias') and automation-distrust bias:

- **Decision-Automation Bias/The Technological Halo Effect:** Users of automated decision-support systems may tend to become hampered in their critical judgment, rational agency, and situational awareness as a result of their faith in the perceived objectivity, neutrality, certainty, or superiority of the AI system.

  This may lead to **over-reliance** or **errors of omission**, where implementers lose the capacity to identify and respond to the faults, errors, or deficiencies, which might arise over the course of the use of an automated system, because they become complacent and overly deferent to its directions and cues. Decision-automation bias may also lead to **over-compliance** or **errors of commission** where implementers defer to the perceived infallibility of the system and thereby become unable to detect problems emerging from its use for reason of a failure to hold the results against available information.

  Both over-reliance and over-compliance may lead to what is known as out-of-loop syndrome where the degradation of the role of human reason and the deskilling of critical thinking hampers the user's ability to complete the tasks that have been automated. This condition may bring about a loss of the ability to respond to system failure and may lead both to safety hazards and to dangers of discriminatory harm.

  To combat risks of decision-automation bias, you should operationalise strong regimes of accountability at the site of user deployment to steer human decision-agents to act on the basis of good reasons, solid inferences, and critical judgment.

- **Automation-Distrust Bias:** At the other extreme, users of an automated decision-support system may tend to disregard its salient contributions to evidence-based reasoning either as a result of their distrust or skepticism about AI technologies in general or as a result of their over-prioritisation of the importance of prudence, common sense, and human expertise. An aversion to the non-human and amoral character of automated systems may also influence decision subjects' hesitation to consult these technologies in high impact contexts such as healthcare, transportation, and law.

In order to secure and safeguard fair implementation of AI systems by users well-trained to utilise the algorithmic outputs as tools for making evidence-based judgements, you should consider the following measures:

- Training of implementers should include the conveyance of basic knowledge about the statistical and probabilistic character of machine learning and about the limitations of AI and automated decision-support technologies. This training should avoid any anthropomorphic (or human-like) portrayals of AI systems and should encourage users to view the benefits and risks of deploying these systems in terms of their role in assisting human judgment rather than replacing it.

- Forethought should be given in the design of the user-system interface about human factors and about possibilities for implementation biases. The systems should be *designed into* processes that encourage active user judgment and situational awareness. The interface between the user and the system should be designed to make clear and accessible to the user the system's rationale, compliance to fairness standards, and confidence level. Ideally this should happen in a 'runtime' manner.

- Training of implementers should include a pre-emptive exploration of the cognitive and judgmental biases that may occur across deployment contexts. This should be done in a use case based manner that highlights the particular misjudgements that may occur when people weigh statistical evidence. Examples of the latter may include overconfidence in prediction based on the historical consistency of data, illusions that any clustering of data points necessarily indicates significant insights, and discounting of societal patterns that exist beyond the statistical results.

### *Putting the principle of discriminatory non-harm into action*

When you are considering how to put the principle of discriminatory non-harm into action, you should come together with all the managers on the project team to map out team member involvement at each stage of the AI project pipeline from alpha through beta. Considering fairness aware design and implementation from a workflow perspective will allow you, as a team, to concretise and make explicit end-to-end paths of accountability in a clear and peer-reviewable manner. This is essential for establishing a robust accountability framework. Here is a schematic representation of the fairness aware workflow. You will have to complete the final row.

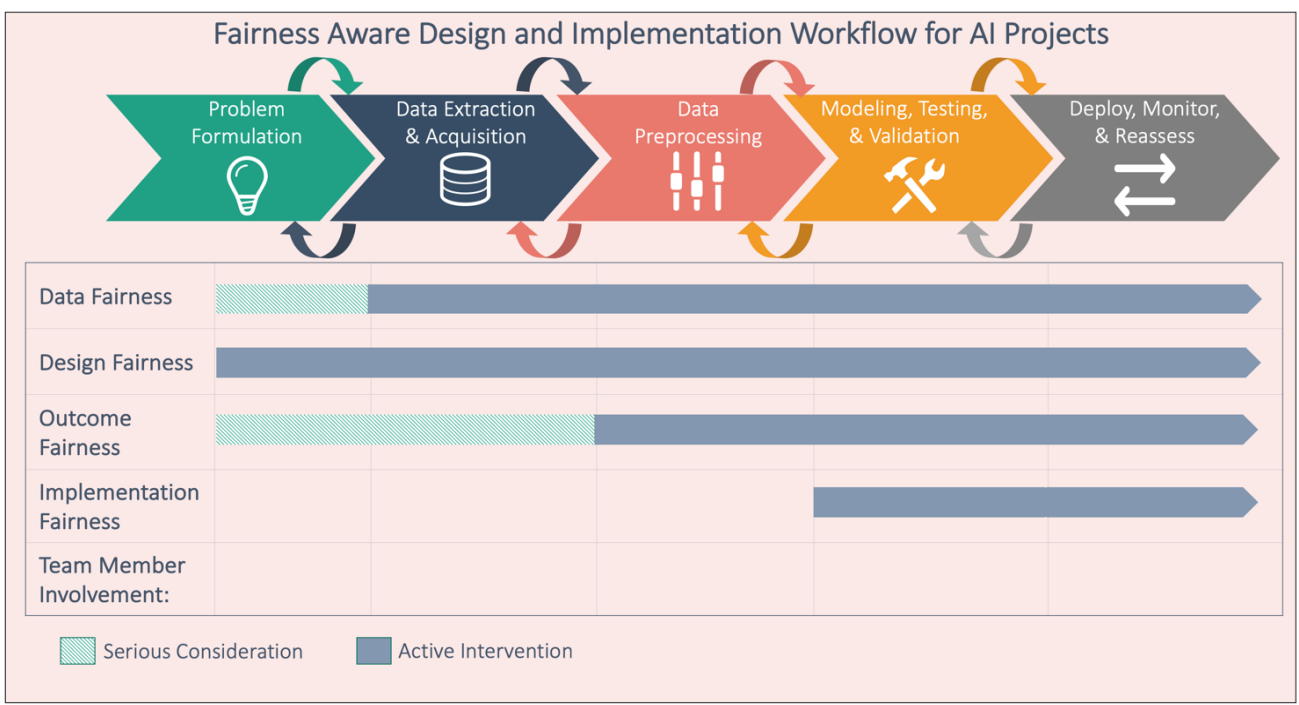

Considering fairness aware design and implementation from such a workflow perspective will also assist you in **pinpointing risks of bias or downstream discrimination and streamlining possible solutions in a proactive, pre-emptive, and anticipatory way.** At each stage of the AI project pipeline (i.e. at each column of the above table), you and the relevant members of your team should carry out a collaborative self-assessment with regard to the applicable dimension of fairness. This is a three-step process:

---

### Discriminatory Non-Harm Self-Assessment

**Step 1:** Identify the fairness and bias mitigation dimensions that apply to the specific stage under consideration (for example, at the data pre-processing stage, dimensions of data fairness, design fairness, and outcome fairness may be at issue).

**Step 2:** Scrutinise how your particular AI project might pose risks or have unintended vulnerabilities in each of these areas.

**Step 3:** Take action to correct any existing problems that have been identified, strengthen areas of weakness that have possible discriminatory consequences, and take proactive bias-prevention measures in areas that have been identified to pose potential future risks.

---

### Accountability

When considering the role of accountability in the AI project delivery lifecycle, it is important first to make sure that you are taking a 'best practices' approach to data processing that is aligned with [Principle 6 of the Data Ethics Framework](#). Beyond following this general guidance, however, you should pay special attention to the new and unique challenges posed to public sector accountability by the design and implementation of AI systems.

Responsible AI project delivery requires that two related challenges to public sector accountability be confronted directly:

1. **Accountability gap:** As mentioned above, automated decisions are not self-justifiable. Whereas human agents can be called to account for their judgements and decisions in instances where those judgments and decisions affect the interests of others, the statistical models and underlying hardware that compose AI systems are not responsible in the same morally relevant sense. This creates an accountability gap that must be addressed so that clear and imputable sources of human answerability can be attached to decisions assisted or produced by an AI system.

2. **Complexity of AI production processes:** Establishing human answerability is not a simple matter when it comes to the design and deployment of AI systems. This is due to the complexity and multi-agent character of the development and use of these systems. Typically, AI project delivery workflows include department and delivery leads, technical experts, data procurement and preparation personnel, policy and domain experts, implementers, and others. Due to this production complexity, it may become difficult to answer the question of who among these parties involved in the production of AI systems should bear responsibility if these systems' uses have negative consequences and impacts.

Meeting the special requirements of accountability, which are born out of these two challenges, call for a sufficiently fine-grained concept of what would make an AI project properly accountable. This concept can be broken down into two subcomponents of accountability: **answerability** and **auditability**:

- **Answerability**: The principle of accountability demands that the onus of justifying algorithmically supported decisions be placed on the shoulders of the human creators and users of those AI systems. This means that it is essential to establish a continuous chain of human responsibility across the whole AI project delivery workflow. Making sure that accountability is effective from end to end necessitates that no gaps be permitted in the answerability of responsible human authorities from first steps of the design of an AI system to its algorithmically steered outcomes.

  Answerability also demands that explanations and justifications of both the content of algorithmically supported decisions and the processes behind their production be offered by competent human authorities in plain, understandable, and coherent language. These explanations and justifications should be based upon sincere, consistent, sound, and impartial reasons that are accessible to non-technical hearers.

- **Auditability**: Whereas the notion of answerability responds to the question of *who is accountable* for an automation supported outcome, the notion of auditability answers the question of *how the designers and implementers of AI systems are to be held accountable*. This aspect of accountability has to do with **demonstrating** both the **responsibility of design and use practices** and the **justifiability of outcomes**.

  Your project team must ensure that every step of the process of designing and implementing your AI project is accessible for audit, oversight, and review. Successful audit requires builders and implementers of algorithmic systems to keep records and to make accessible information that enables monitoring of the soundness and diligence of the innovation processes that produced the AI system.

  Auditability also requires that your project team keep records and make accessible information that enables monitoring of data provenance and analysis from the stages of collection, pre-processing, and modelling to training, testing, and deploying. This is the purpose of the previously mentioned Dataset Factsheet.

  Moreover, it requires your team to enable peers and overseers to probe and to critically review the dynamic operation of the system in order to ensure that the procedures and operations which are producing the model's behaviour are safe, ethical, and fair. Practically transparent algorithmic models must be **built for auditability**, **reproducible**, and **equipped for end-to-end recording and monitoring** of their data processing.

The deliberate incorporation of both of these elements of accountability (answerability and auditability) into the AI project lifecycle may be called **Accountability-by-Design**:

> **Accountability by Design:** All AI systems must be designed to facilitate end-to-end answerability and auditability. This requires both *responsible humans-in-the-loop* across the entire design and implementation chain as well as *activity monitoring protocols* that enable end-to-end oversight and review.

*Accountability deserves consideration across the entire design and implementation workflow*

As a best practice, you should actively consider the different demands that accountability by design places on you before and after the roll out of your AI project. We will refer to the process of ensuring accountability during the design and development stages of your AI project as '**anticipatory accountability**.' This is because you are anticipating your AI project's accountability needs prior to it being completed. Following a similar logic, we will refer to the process of addressing accountability after the start of the deployment of your AI project as '**remedial accountability**.' This is because after the initial implementation of your system, you are remedying any of the issues that may be raised by its effects and potential externalities. These two subtypes of accountability are sometimes referred to as *ex-ante* (or before-the-event) accountability and *ex-post* (after-the-event) accountability respectively.

- **Anticipatory Accountability:** Treating accountability as an anticipatory principle entails that you take as of primary importance the decisions made and actions taken by your project delivery team prior to the outcome of an algorithmically supported decision process.

  This kind of *ex ante* accountability should be prioritised over remedial accountability, which focuses instead on the corrective or justificatory measures that can be taken after that automation supported process had been completed.

  By ensuring the AI project delivery processes are accountable prior to the actual application of the system in the world, you will bolster the soundness of design and implementation processes and thereby more effectively pre-empt possible harms to individual wellbeing and public welfare.

  Likewise, by establishing strong regimes of anticipatory accountability and by making the design and delivery process as open and publicly accessible as possible, you will put affected stakeholders in a position to make better informed and more knowledgeable decisions about their involvement with these systems in advance of potentially harmful impacts. In doing so, you will also strengthen the public narrative and help to safeguard the project from reputational harm.

- **Remedial Accountability:** While remedial accountability should be seen, along these lines, as a necessary fallback rather than as a first resort for imputing responsibility in the design and deployment of AI systems, strong regimes of remedial accountability are no less important in providing necessary justifications for the bearing these systems have on the lives of affected stakeholders.

**Putting in place comprehensive auditability regimes as part of your accountability framework and establishing transparent design and use practices, which are methodically logged throughout the AI project delivery lifecycle, are essential components for this sort of remedial accountability.**

One aspect of remedial accountability that you must pay close attention to is the need to provide **explanations** to affected stakeholders for algorithmically supported decisions. This aspect of accountable and transparent design and use practices will be called **explicability,** which literally means the ability to make explicit the meaning of the algorithmic model's result.

Offering explanations for the results of algorithmically supported decision-making involves furnishing decision subjects and other interested parties with an understandable account of the rationale behind the specific outcome of interest. It also involves furnishing the decision subject and other interested parties with an explanation of the ethical permissibility, the fairness, and the safety of the use of the AI system. These tasks of **content clarification** and **practical justification** will be explored in more detail below as part of the section on transparency.

## Sustainability

Designers and users of AI systems should remain aware that these technologies may have transformative and long-term effects on individuals and society. In order to ensure that the deployment of your AI system remains sustainable and supports the sustainability of the communities it will affect, you and your team should proceed with a continuous sensitivity to the real-world impacts that your system will have.

### *Stakeholder Impact Assessment*

You and your project team should come together to evaluate the social impact and sustainability of your AI project through a **Stakeholder Impact Assessment (SIA)**, whether the AI project is being used to deliver a public service or in a back-office administrative capacity. When we refer to 'stakeholders' we are referring primarily to affected individual persons, but the term may also extend to groups and organisations in the sense that individual members of these collectives may also be impacted as such by the design and deployment of AI systems. Due consideration to stakeholders should be given at both of these levels.

The purpose of carrying out an SIA is multidimensional. SIAs can serve several purposes, some of which include:

(1) Help to build public confidence that the design and deployment of the AI system by the public sector agency has been done responsibly
(2) Facilitate and strengthen your accountability framework
(3) Bring to light unseen risks that threaten to affect individuals and the public good

    (4) Underwrite well-informed decision-making and transparent innovation practices

    (5) Demonstrate forethought and due diligence not only within your organisation but also to the wider public

Your team should convene to evaluate the social impact and sustainability of your AI project through the SIA at three critical points in the project delivery lifecycle:

1. **Alpha Phase (Problem Formulation):** Carry out an initial Stakeholder Impact Assessment (SIA) to determine the ethical permissibility of the project. Refer to the SUM Values as a starting point for the considerations of the possible effects of your project on individual wellbeing and public welfare. In cases where you conclude that your AI project will have significant ethical and social impacts, you should open your initial SIA to the public so that their views can be properly considered. This will bolster the inclusion of a diversity of voices and opinions into the design and development process through the participation of a more representative range of stakeholders. You should also consider consulting with internal organisational stakeholders, whose input will likewise strengthen the openness, inclusivity, and diversity of your project.

2. **From Alpha to Beta (Pre-Implementation):** Once your model has been trained, tested, and validated, you and your team should revisit your initial SIA to confirm that the AI system to be implemented is still in line with the evaluations and conclusions of your original assessment. This check-in should be logged on the pre-implementation section of the SIA with any applicable changes added and discussed. Before the launch of the system, this SIA should be made publicly available. At this point you must also set a timeframe for re-assessment once the system is in operation as well as a public consultation which predates and provides input for that re-assessment. Timeframes for these re-assessments should be decided by your team on a case-by-case basis but should be proportional to the scale of the potential impact of the system on the individuals and communities it will affect.

3. **Beta Phase (Re-Assessment):** After your AI system has gone live, your team should intermittently revisit and re-evaluate your SIA. These check-ins should be logged on the re-assessment section of the SIA with any applicable changes added and discussed. Re-assessment should focus both on evaluating the existing SIA against real world impacts and on considering how to mitigate the unintended consequences that may have ensued in the wake of the deployment of the system. Further public consultation for input at the beta stage should be undertaken before the re-assessment so that stakeholder input can be included in re-assessment deliberations.

You should keep in mind that, in its specific focus on social and ethical sustainability, your Stakeholder Impact Assessment constitutes just one part of the governance platform for your AI project and should be a complement to your accountability framework and other auditing and activity-monitoring documentation.

Your SIA should be broken down into four sections of questions and responses. In the 1st section, there should be general questions about the possible big-picture social and ethical impacts of the use of the AI system you plan to build. In the 2nd section, your team should collaboratively formulate relevant sector-specific and use case-specific questions about the impact of the AI system on affected stakeholders. The 3<sup>rd</sup> section should provide answers to the additional questions relevant to pre-implementation evaluation. The 4<sup>th</sup> section should provide the opportunity for members of your team to reassess the system in light of its real-world impacts, public input, and possible unintended consequences.

Here is a prototype of an SIA:

| Stakeholder Impact Assessment for (Project Name) | |
|---|---|
| 1. Alpha Phase (Problem Formulation) General Questions<br><br>Completed on this Date: | **I. Identifying Affected Stakeholders**<br><br>Who are the stakeholders that this AI project is most likely to affect? What groups of these stakeholders are most vulnerable? How might the project negatively impact them?<br><br>**II. Goal-Setting and Objective-Mapping**<br><br>How are you defining the outcome (the target variable) that the system is optimising for? Is this a fair, reasonable, and widely acceptable definition?<br><br>Does the target variable (or its measurable proxy) reflect a reasonable and justifiable translation of the project's objective into the statistical frame?<br><br>Is this translation justifiable given the general purpose of the project and the potential impacts that the outcomes of its implementation will have on the communities involved?<br><br>**III. Possible Impacts on the Individual**<br><br>How might the implementation of your AI system impact the abilities of affected stakeholders to make free, independent, and well-informed decisions about their lives? How might it enhance or diminish their autonomy?<br><br>How might it affect their capacities to flourish and to fully develop themselves?<br><br>How might it do harm to their physical or mental integrity? Have risks to individual health and safety been adequately considered and addressed?<br><br>How might it infringe on their privacy rights, both on the data processing end of designing the system and on the implementation end of deploying it?<br><br>**IV. Possible Impacts on Society and Interpersonal Relationships**<br><br>How might the implementation of your AI system adversely affect each stakeholder's fair and equal treatment under the law? Are there any aspects of the project that expose vulnerable communities to possible discriminatory harm?<br><br>How might the use of your system affect the integrity of interpersonal dialogue, meaningful human connection, and social cohesion? |

| | Have the values of civic participation, inclusion, and diversity been adequately considered in articulating the purpose and setting the goals of the project? If not, how might these values be incorporated into your project design? |
|---|---|
| | Does the project aim to advance the interests and well-being of as many affected individuals as possible? Might any disparate socioeconomic impacts result from its deployment? |
| | Have you sufficiently considered the wider impacts of the system on future generations and on the planet as a whole? |
| 2. Alpha Phase (Problem Formulation) Sector-Specific and Use Case-Specific Questions<br><br>Completed on this Date: | In this section you should consider the sector-specific and use case-specific issues surrounding the social and ethical impacts of your AI project on affected stakeholders. Compile a list of the questions and concerns you anticipate. State how your team is attempting to address these questions and concerns. |
| 3. From Alpha to Beta (Pre-Implementation)<br><br>Completed on this Date: | After reviewing the results of your initial SIA, answer the following questions:<br><br>Are the trained model's actual objective, design, and testing results still in line with the evaluations and conclusions contained in your original assessment? If not, how does your assessment now differ?<br><br>Have any other areas of concern arisen with regard to possibly harmful social or ethical impacts as you have moved from the alpha to the beta phase?<br><br>You must also set a reasonable timeframe for public consultation and beta phase re-assessment:<br><br>**Dates of Public Consultation on Beta-Phase Impacts:**<br><br>**Date of Planned Beta Phase Re-Assessment:** |
| 4. Beta Phase (Re-Assessment)<br><br>Completed on this Date: | Once you have reviewed the most recent version of your SIA and the results of the public consultation, answer the following questions:<br><br>How does the content of the existing SIA compare with the real-world impacts of the AI system as measured by available evidence of performance, monitoring data, and input from implementers and the public?<br><br>What steps can be taken to rectify any problems or issues that have emerged?<br><br>Have any unintended harmful consequences ensued in the wake of the deployment of the system? If so, how might these negative impacts be mitigated and redressed? |

| | Have the maintenance processes for your AI model adequately taken into account the possibility of distributional shifts in the underlying population? Has the model been properly retuned and retrained to accommodate changes in the environment?<br><br>Dates of Public Consultation on Beta-Phase Impacts:<br><br>Date of Next Planned Beta Phase Re-Assessment: |
|---|---|

## Safety

Beyond safeguarding the sustainability of your AI project as it relates to its social impacts on individual wellbeing and public welfare, your project team must also confront the related challenge of **technical sustainability** or **safety**. A technically sustainable AI system is **safe, accurate, reliable, secure, and robust**. Securing these goals, however, is a difficult and unremitting task.

Because AI systems operate in a world filled with uncertainty, volatility, and flux, the challenge of building safe and reliable AI can be especially daunting. This job, however, must be met head-on. Only by making the goal of producing safe and reliable AI technologies central to your project, will you be able to mitigate risks of your system failing at scale when faced with real-world unknowns and unforeseen events. The issue of **AI safety** is of paramount importance, because these potential failures may both produce harmful outcomes and undermine public trust.

In order to safeguard that your AI system functions safely, you must prioritise the technical objectives of **accuracy, reliability, security, and robustness**. This requires that your technical team put careful forethought into how to construct a system **that accurately and dependably operates in accordance with its designers' expectations even when confronted with unexpected changes, anomalies, and perturbations**. Building an AI system that meets these safety goals also requires rigorous testing, validation, and re-assessment as well as the integration of adequate mechanisms of oversight and control into its real-world operation.

### *Accuracy, reliability, security, and robustness*

It is important that you gain a strong working knowledge of each of the safety relevant operational objectives (**accuracy, reliability, security, and robustness**):

- **Accuracy and Performance Metrics:** In machine learning, the accuracy of a model is the proportion of examples for which it generates a correct output. This performance measure is also sometimes characterised conversely as an **error rate** or the fraction of cases for which the model produces an incorrect output. Keep in mind that, in some instances, the choice of an acceptable error rate or accuracy level can be adjusted in accordance with the use case specific needs of the application. In other instances, it may be largely set by a domain established benchmark.

As a performance metric, accuracy should be a central component to establishing and nuancing your team's approach to safe AI. That said, specifying a reasonable performance level for your system may also often require you to refine or exchange your measure of accuracy. For instance, if certain errors are more significant or costly than others, a metric for total cost can be integrated into your model so that the cost of one class of errors can be weighed against that of another. Likewise, if the precision and sensitivity of the system in detecting uncommon events is a priority (say, in instances of the medical diagnosis of rare cases of a disease), you can use the technique of precision and recall. This method of addressing imbalanced classification would allow you to weigh the proportion of the system's correct detections—both of frequent and of rare outcomes—against the proportion of actual detections of the rare event (i.e. the ratio of the true detections of the rare outcome to the sum of the true detections of that outcome and the missed detections or false negatives for that outcome).

In general, measuring accuracy in the face of uncertainty is a challenge that must be given significant thought. The confidence level of your AI system will depend heavily on problems inherent in attempts to model a chaotic and changing reality. Concerns about accuracy must cope with issues of unavoidable noise present in the data sample, architectural uncertainties generated by the possibility that a given model is missing relevant features of the underlying distribution, and inevitable changes in input data over time.

- **Reliability:** The objective of reliability is that an AI system behaves exactly as its designers intended and anticipated. A reliable system adheres to the specifications it was programmed to carry out.  Reliability is therefore a measure of **consistency** and can establish confidence in the safety of a system based upon the dependability with which it operationally conforms to its intended functionality.

- **Security:** The goal of security encompasses the protection of several operational dimensions of an AI system when confronted with possible adversarial attack. A secure system is capable of maintaining the **integrity** of the information that constitutes it. This includes protecting its architecture from the unauthorised modification or damage of any of its component parts. A secure system also remains continuously **functional** and **accessible** to its authorised users and keeps **confidential** and **private information** secure even under hostile or adversarial conditions.

- **Robustness:**  The objective of robustness can be thought of as the goal that an AI system functions reliably and accurately under harsh conditions. These conditions may include adversarial intervention, implementer error, or skewed goal-execution by an automated learner (in reinforcement learning applications). The measure of robustness is therefore the strength of a system's integrity and the soundness of its operation in response to difficult conditions, adversarial attacks, perturbations, data poisoning, and undesirable reinforcement learning behaviour.

*Risks posed to accuracy and reliability:*

**Concept Drift**: Once trained, most machine learning systems operate on static models of the world that have been built from historical data which have become fixed in the systems' parameters. This freezing of the model before it is released 'into the wild' makes its accuracy and reliability especially vulnerable to changes in the underlying distribution of data. When the historical data that have crystallised into the trained model's architecture cease to reflect the population concerned, the model's mapping function will no longer be able to accurately and reliably transform its inputs into its target output values. These systems can quickly become prone to error in unexpected and harmful ways.

There has been much valuable research done on methods of detecting and mitigating concept and distribution drift, and you should consult with your technical team to ensure that its members have familiarised themselves with this research and have sufficient knowledge of the available ways to confront the issue. In all cases, you should remain vigilant to the potentially rapid concept drifts that may occur in the complex, dynamic, and evolving environments in which your AI project will intervene. Remaining aware of these transformations in the data is crucial for safe AI, and your team should actively formulate an action plan to anticipate and to mitigate their impacts on the performance of your system.

**Brittleness:** Another possible challenge to the accuracy and reliability of machine learning systems arises from the inherent imitations of the systems themselves. Many of the high-performing machine learning models such as deep neural nets (DNN) rely on massive amounts of data and brute force repetition of training examples to tune the thousands, millions, or even billions of parameters, which collectively generate their outputs.

However, when they are actually running in an unpredictable world, these systems may have difficulty processing unfamiliar events and scenarios. They may make unexpected and serious mistakes, because they have neither the capacity to contextualise the problems they are programmed to solve nor the common-sense ability to determine the relevance of new 'unknowns'. Moreover, these mistakes may remain unexplainable given the high-dimensionality and computational complexity of their mathematical structures.  This fragility or brittleness can have especially significant consequences in safety-critical applications like fully automated transportation and medical decision support systems where undetectable changes in inputs may lead to significant failures. While progress is being made in finding ways to make these models more robust, it is crucial to consider safety first when weighing up their viability.

*Risks posed to security and robustness*

**Adversarial Attack:** Adversarial attacks on machine learning models maliciously modify input data—often in imperceptible ways—to induce them into misclassification or incorrect prediction. For instance, by undetectably altering a few pixels on a picture, an adversarial attacker can mislead a model into generating an incorrect output (like identifying a panda as a gibbon or a 'stop' sign as a 'speed limit' sign) with an extremely high confidence. While a good amount of attention has been paid to the risks that adversarial attacks pose in deep learning applications like computer vision, these kinds of perturbations are also effective across a vast range of machine learning techniques and uses such as spam filtering and malware detection.

These vulnerabilities of AI systems to adversarial examples have serious consequences for AI safety. The existence of cases where subtle but targeted perturbations cause models to be misled into gross miscalculation and incorrect decisions have potentially serious safety implication for the adoption of critical systems like applications in autonomous transportation, medical imaging, and security and surveillance. In response to concerns about the threats posed to a safe and trusted environment for AI technologies by adversarial attacks a field called **adversarial machine learning** has emerged over the past several years. Work in this area focuses on securing systems from disruptive perturbations at all points of vulnerability across the AI pipeline.

One of the major safety strategies that has arisen from this research is an approach called **model hardening**, which has advanced techniques that combat adversarial attacks by strengthening the architectural components of the systems. Model hardening techniques may include adversarial training, where training data is methodically enlarged to include adversarial examples. Other model hardening methods involve architectural modification, regularisation, and data pre-processing manipulation. A second notable safety strategy is **run-time detection**, where the system is augmented with a discovery apparatus that can identify and trace in real-time the existence of adversarial examples. You should consult with members of your technical team to ensure that the risks of adversarial attack have been taken into account and mitigated throughout the AI lifecycle. A valuable collection of resources to combat adversarial attack can be found at https://github.com/IBM/adversarial-robustness-toolbox .

**Data Poisoning:** A different but related type of adversarial attack is called data poisoning. This threat to safe and reliable AI involves a malicious compromise of data sources at the point of collection and pre-processing. Data poisoning occurs when an adversary modifies or manipulates part of the dataset upon which a model will be trained, validated, and tested. By altering a selected subset of training inputs, a poisoning attack can induce a trained AI system into curated misclassification, systemic malfunction, and poor performance. An especially concerning dimension of targeted data poisoning is that an adversary may introduce a 'backdoor' into the infected model whereby the trained system functions normally until it processes maliciously selected inputs that trigger error or failure.

In order to combat data poisoning, your technical team should become familiar with the state of the art in filtering and detecting poisoned data. However, such technical solutions are not enough. Data poisoning is possible because data collection and procurement often involves potentially unreliable or questionable sources. When data originates in uncontrollable environments like the internet, social media, or the Internet of Things, many opportunities present themselves to ill-intentioned attackers, who aim to manipulate training examples. Likewise, in third- party data curation processes (such as 'crowdsourced' labelling, annotation, and content identification), attackers may simply handcraft malicious inputs. Your project team should focus on the best practices of responsible data management, so that they are able to tend to data quality as an end-to-end priority.

- **Misdirected Reinforcement Learning Behaviour:** A different set of safety risks emerges from the approach to machine learning called reinforcement learning (RL). In the more widely applied methods of supervised learning that have largely been the focus of this guide, a model transforms inputs into outputs according to a fixed mapping function that has resulted from its passively received training. In RL, by contrast, the learner system actively solves problems by engaging with its environment through trial and error. This exploration and 'problem-solving' behaviour is determined by the objective of maximising a reward function that is defined by its designers.

This flexibility in the model, however, comes at the price of potential safety risks. An RL system, which is operating in the real-world without sufficient controls, may determine a reward-optimising course of action that is optimal for achieving its desired objective but harmful to people. Because these models lack context-awareness, common sense, empathy, and understanding, they are unable to identify, on their own, scenarios that may have damaging consequences but that were not anticipated and constrained by their programmers. This is a difficult problem, because the unbounded complexity of the world makes anticipating all of its pitfalls and detrimental variables veritably impossible.

Existing strategies to mitigate such risks of misdirected reinforcement learning behaviour include:
- o Running extensive simulations during the testing stage, so that appropriate measures of constraint can be programmed into the system
- o Continuous inspection and monitoring of the system, so that its behaviour can be better predicted and understood
- o Finding ways to make the system more interpretable so that its decisions can be better assessed
- o Hard-wiring mechanisms into the system that enable human override and system shut-down

## End-to-End AI Safety

The safety risks you face in your AI project will depend, among other factors, on the sort of algorithm(s) and machine learning techniques you are using, the type of applications in which those techniques are going to be deployed, the provenance of your data, the way you are specifying your objective, and the problem domain in which that specification applies. As a best practice, regardless of this variability of techniques and circumstances, safety considerations of accuracy, reliability, security, and robustness should be in operation at every stage of your AI project lifecycle.

This should involve both **rigorous protocols of testing, validating, verifying, and monitoring the safety of the system** and the performance of **AI safety self-assessments** by relevant members of your team at each stage of the workflow. Such self-assessments should evaluate how your team's design and implementation practices line up with the safety objectives of accuracy, reliability, security, and robustness. Your AI safety self-assessments should be logged across the workflow on a single document in a running fashion that allows review and re-assessment.

## Transparency

## Defining transparent AI

It is important to remember that ***transparency as a principle of AI ethics*** differs a bit in meaning from the everyday use of the term. The common dictionary understanding of transparency defines it as *either* (1) the quality an object has when one can see clearly through it or (2) the quality of a situation or process that can be clearly justified and explained because it is open to inspection and free from secrets.

Transparency as a principle of AI ethics encompasses *both* of these meanings:

On the one hand, transparent AI involves the interpretability of a given AI system, i.e. **the ability to know how and why a model performed the way it did in a specific context and therefore to understand the rationale behind its decision or behaviour**. This sort of transparency is often referred to by way of the metaphor of 'opening the black box' of AI. It involves ***content clarification and intelligibility*** or **explicability.**

On the other hand, transparent AI involves **the justifiability both of the processes that go into its design and implementation and of its outcome**. It therefore involves the ***soundness of the justification of its use***. In this more normative meaning, transparent AI is ***practically justifiable*** in an unrestricted way if one can demonstrate that both the design and implementation processes that have gone into the particular decision or behaviour of a system and the decision or behaviour itself are **ethically permissible, non-discriminatory/fair, and worthy of public trust/safety-securing.**

### *Three critical tasks for designing and implementing transparent AI*

This two-pronged definition of transparency as a principle of AI ethics asks that you to think about transparent AI both in terms of the *process* behind it (the design and implementation practices that lead to an algorithmically supported outcome) and in terms of its *product* (the content and justification of that outcome).  Such a process/product distinction is crucial, because it clarifies the three tasks that your team will be responsible for in safeguarding the transparency of your AI project:

- **Process Transparency, Task 1: Justify Process.** In offering an explanation to affected stakeholders, you should be able to demonstrate that considerations of ethical permissibility, non-discrimination/fairness, and safety/public trustworthiness were operative end-to-end in the design and implementation processes that lead to an automated decision or behaviour. This task will be supported both by following the best practices outlined herein throughout the AI project lifecycle and by putting into place robust auditability measures through an accountability-by-design framework.

- **Outcome Transparency, Task 2: Clarify Content and Explain Outcome.** In offering an explanation to affected stakeholders, you should be able to show in plain language that is understandable to non-specialists how and why a model performed the way it did in a specific decision-making or behavioural context. You should therefore be able to clarify and communicate the rationale behind its decision or behaviour. This explanation should be *socially meaningful* in the sense that the terms and logic of the explanation should not simply reproduce the formal characteristics or the technical meanings and rationale of the mathematical model but should rather be translated into the everyday language of human practices and therefore be understandable in terms of the societal factors and relationships that the decision or behaviour implicates.

- **Outcome Transparency, Task 3: Justify Outcome.** In offering an explanation to affected stakeholders, you should be able to demonstrate that a specific decision or behaviour of your system is ethically permissible, non-discriminatory/fair, and worthy of public trust/safety-securing. This outcome justification should take the content clarification/explicated outcome from task 2 as its starting point and weigh that explanation against the justifiability criteria adhered to throughout the design and use pipeline: ethical permissibility, non-discrimination/fairness, and safety/public trustworthiness. Undertaking an optimal approach to process transparency from the start should support and safeguard this demand for normative explanation and outcome justification.

*Mapping AI transparency*

Before exploring each of the three tasks individually, it may be helpful to visualise the relationship between these connected components of transparent AI:

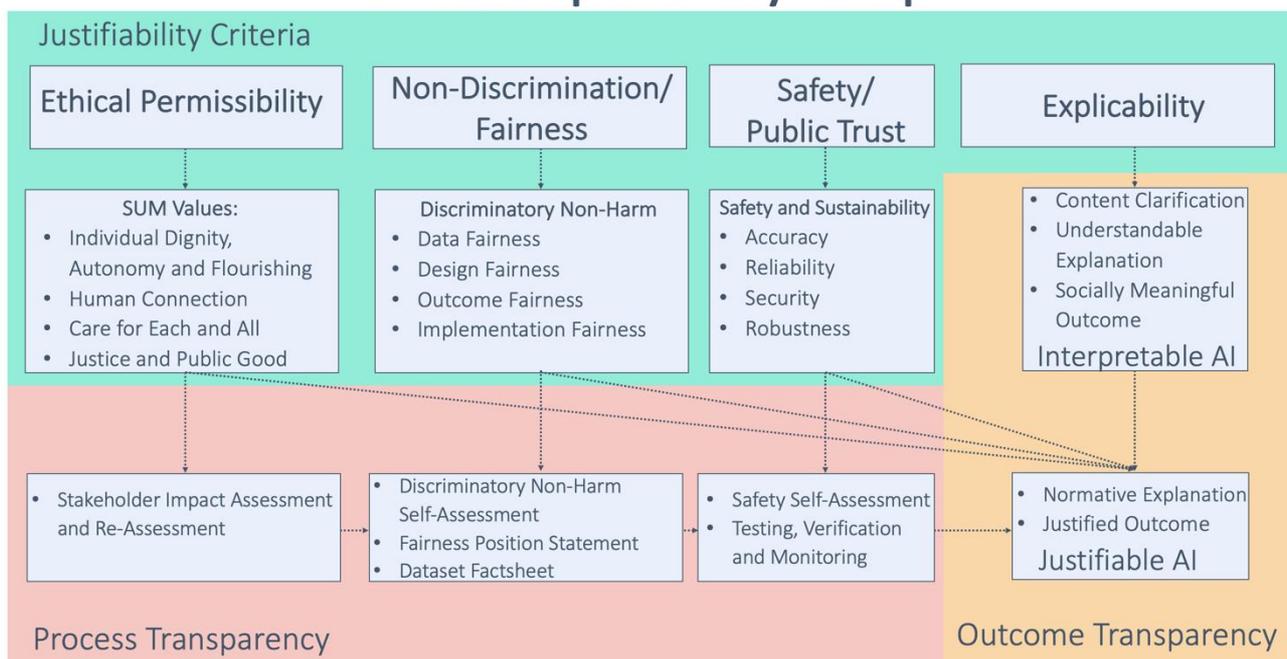

# Process Transparency: Establishing a Process-Based Governance Framework

The central importance of the end-to-end operability of good governance practices should guide your strategy to build out responsible AI project workflow processes. Three components are essential to creating a such a responsible workflow: (1) Maintaining strong regimes of professional and institutional transparency; (2) Having a clear and accessible Process-Based Governance

Framework (PBG Framework); (3) Establishing a well-defined auditability trail in your PBG Framework through robust activity logging protocols that are consolidated digitally in a process log.

1. **Professional and Institutional Transparency:** At every stage of the design and implementation of your AI project, team members should be held to rigorous standards of conduct that secure and maintain professionalism and institutional transparency. These standards should include the core values of **integrity, honesty, sincerity, neutrality, objectivity and impartiality**. All professionals involved in the research, development, production, and implementation of AI technologies are, first and foremost, acting as **fiduciaries of the public interest** and must, in keeping with these core values of the Civil Service, put the obligations to serve that interest above any other concerns.

   Furthermore, from start to finish of the AI project lifecycle, the design and implementation process should be as transparent and as open to public scrutiny as possible with restrictions on accessibility to relevant information limited to the reasonable protection of justified public sector confidentiality and of analytics that may tip off bad actors to methods of gaming the system of service provision.

2. **Process-Based Governance Framework:** So far, this guide has presented some of the main steps that are necessary for establishing responsible innovation practices in your AI project. Perhaps the most vital of these measures is the effective operationalisation of the values and principles that underpin the development of ethical and safe AI. By organising all of your governance considerations and actions into a PBG Framework, you will be better able to accomplish this task.

   The purpose of a PBG Framework is to provide a template for the integrations of the norms, values, and principles, which motivate and steer responsible innovation, with the actual processes that characterise the AI design and development pipeline. While the accompanying Guide has focused primarily on the Cross Industry Standard Process for Data Mining (CRISP-DM), keep in mind that such a structured integration of values and principles with innovation processes is just as applicable in other related workflow models like Knowledge Discovery in Databases (KDD) and Sample, Explore, Modify, Model, and Assess (SEMMA).

   Your PBG Framework should give you a landscape view of the governance procedures and protocols that are organising the control structures of your project workflow. Constructing a good PBG Framework will provide you and your team with a big picture of:

   - The relevant team members and roles involved in each governance action.
   - The relevant stages of the workflow in which intervention and targeted consideration are necessary to meet governance goals
   - Explicit timeframes for any necessary follow-up actions, re-assessments, and continual monitoring
   - Clear and well-defined protocols for logging activity and for instituting mechanisms to assure end-to-end auditability

To help you get a summary picture of where the components of process transparency explored so far fit into a PBG Framework, here is a landscape view:

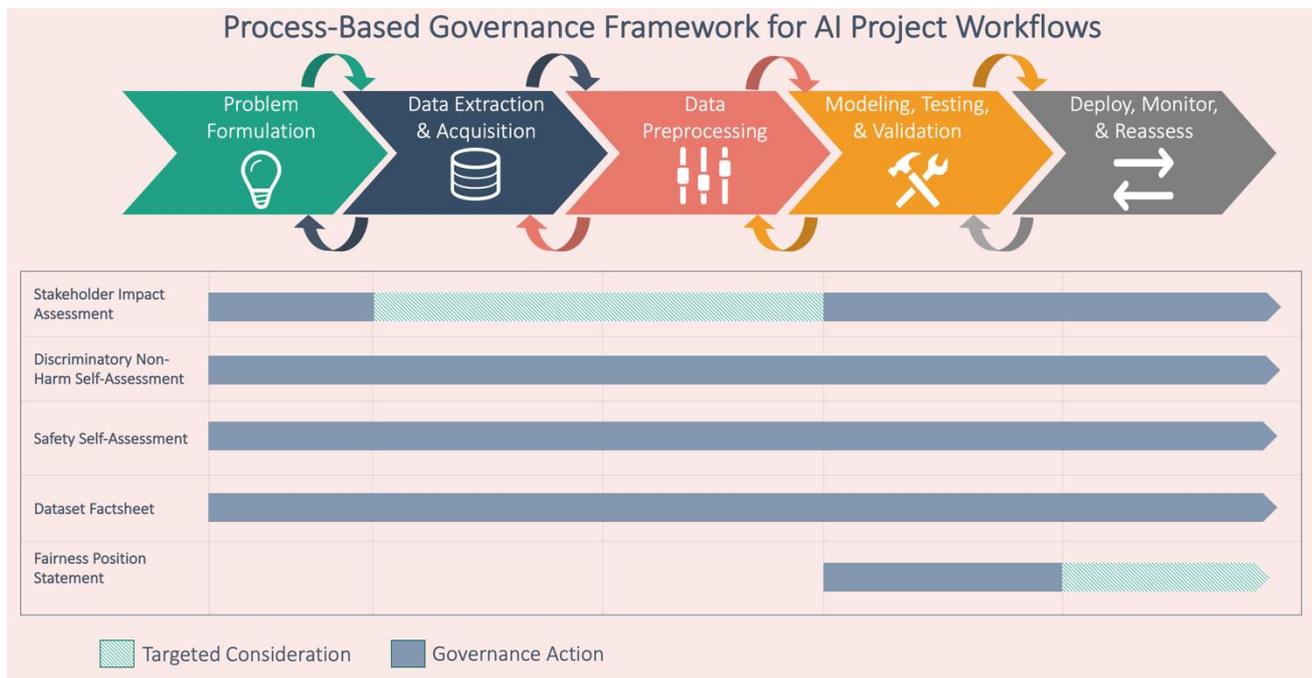

3. **Enabling Auditability with a Process Log:** With your controls in place and your governance framework organised, you will be better able to manage and consolidate the information necessary to assure end-to-end auditability. This information should include both the records and activity-monitoring results that are yielded by your PBG Framework and the model development data gathered across the modelling, training, testing, verifying, and implementation phases.

By centralising your information digitally in a process log, you are preparing the way for optimal process transparency. A process log will enable you to make available, in one place, information that may assist you in demonstrating to concerned parties and affected decision subjects both the responsibility of design and use practices and the justifiability of the outcomes of your system's processing behaviour.

Such a log will also allow you to differentially organise the accessibility and presentation of the information yielded by your project. Not only is this crucial to preserving and protecting data that legitimately should remain unavailable for public view, it will afford your team the capacity to cater the presentation of results to different tiers of stakeholders with different interests and levels of expertise. This ability to curate your explanations with the user-receiver in mind will be vital to achieving the goals of interpretable and justifiable AI.

## Outcome transparency: Explaining outcome and clarifying content

Beyond enabling process transparency through your PBG Framework, you must also put in place standards and protocols to ensure that clear and understandable explanations of the outcomes of your AI system's decisions, behaviours, and problem-solving tasks can:

1. Properly inform the evidence-based judgments of the implementers that they are designed to support;

2. Be offered to affected stakeholders and concerned parties in an accessible way.

This is a multifaceted undertaking that will demand careful forethought and participation across your entire project team.

There is no simple technological solution for how to effectively clarify and convey the rationale behind a model's output in a particular decision-making or behavioural context. Your team will have to use sound judgement and common sense in order to bring together the **technical aspects** of choosing, designing, using a sufficiently interpretable AI system and the **delivery aspects** of being able to clarify and communicate in plain, non-technical, and socially meaningful language how and why that system performed the way it did in a specific decision-making or behavioural context.

Having a good grasp of the rationale and criteria behind the decision-making and problem-solving behaviour of your system is essential for producing safe, fair, and ethical AI. If your AI model is not sufficiently interpretable—if you aren't able to draw from it humanly understandable explanations of the factors that played a significant role in determining its behaviours—then you may not be able to tell how and why things go wrong in your system when they do.

This is a crucial and unavoidable issue for reasons we have already explored. Ensuring the safety of high impact systems in transportation, medicine, infrastructure, and security requires human verification that these systems have properly learned the critical tasks they are charged to complete. It also requires confirmation that when confronted with unfamiliar circumstances, anomalies, and perturbations, these systems will not fail or make unintuitive errors. Moreover, ensuring that these systems operate without causing discriminatory harms requires effective ways to detect and to mitigate sources of bias and inequitable influence that may be buried deep within their feature spaces, inferences, and architectures. Without interpretability each one of these tasks necessary for delivering safe and morally justifiable AI will remain incomplete.

## Defining Interpretable AI

To gain a foothold in both the technical and delivery dimensions of AI interpretability, you will first need a solid working definition of what interpretable AI is. To this end, it may be useful to recall once again the definition of AI offered in the accompanying Guide: 'Artificial Intelligence is the science of *making computers do things that require intelligence when done by humans*.'

This characterisation is important, because it brings out an essential feature of the explanatory demands of interpretable AI: to do things that require intelligence when done by humans means to do things that require *reasoning processes and cognitive functioning*. This cognitive dimension has a direct bearing on how you should think about offering suitable explanations about algorithmically generated outcomes:

**Explaining an algorithmic model's decision or behaviour should involve making explicit how the particular set of factors which determined that outcome can play the role of evidence in supporting**

the conclusion reached. It should involve making intelligible to affected individuals the rationale behind that decision or behaviour as if it had been produced by a reasoning, evidence-using, and inference-making person.

What makes this explanation-giving task so demanding when it comes to AI systems is that reasoning processes do not occur, for humans, at just one level. Rather, human-scale reasoning and interpreting includes:

1. Aspects of **logic** (applying the basic principles of validity that lie behind and give form to sound thinking): *This aspect aligns with the need for **formal or logical explanations** of AI systems.*

2. Aspects of **semantics** (gaining an understanding of how and why things work the way they do and what they mean): *This aspect aligns with the need for **explanations of the technical rationale** behind the outcomes AI systems.*

3. Aspects of the **social understanding of practices, beliefs, and intentions** (clarifying the content of interpersonal relations, societal norms, and individual objectives): *This aspect aligns with the need for the **clarification of the socially meaningful content** of the outcomes of AI systems.*

4. Aspects of **moral justification** (making sense of what should be considered right and wrong in our everyday activities and choices): *This aspect aligns with the **justifiability** of AI systems.*

There are good reasons why ***all four of these dimensions of human reasoning processes*** must factor in to explaining the decisions and behaviours of AI systems: First and most evidently, understanding the logic and technical innerworkings (i.e. semantic content) of these systems is a precondition for ensuring their safety and fairness. Secondly, because they are designed and used to achieve human objectives and to fulfil surrogate cognitive functions *in the everyday social world*, we need to make sense of these systems in terms of the consequential roles that their decisions and behaviours play in that human reality. The social context of these outcomes matters greatly. Finally, because they actually affect individuals and society in direct and morally consequential ways, we need to be able to understand and explain their outcomes not just in terms of their mathematical logic, technical rationale, and social context but also in terms of the justifiability of their impacts on people.

Delving more deeply into the technical and delivery aspects of interpretable AI will show how these four dimensions of human reasoning directly line up with the different levels of demand for explanations of the outcomes of AI systems. In particular, the logical and semantic dimensions will weigh heavily in technical considerations whereas the social and moral dimensions will be significant at the point of delivery.

Note here, though, that these different dimensions of human reasoning are not necessarily mutually exclusive but build off and depend upon each other in significant and cascading ways. Approaching explanations of interpretable AI should therefore be treated holistically and inclusively. Technical explanation of the logic and rationale of a given model, for instance, should be seen as a support for the context-based clarification of its socially meaningful content, just as that socially meaningful content should be viewed as forming the basis of explaining an outcome's moral justifiability. When considering how to make the outcomes of decision-making and problem-solving AI systems maximally transparent to affected stakeholders, you should take this rounded view of human reasoning into account, because it will help you address more effectively the spectrum of concerns that these stakeholders may have.

## Technical aspects of choosing, designing, and using an interpretable AI system

Keep in mind that, while, on the face of it, the task of choosing between the numerous AI and machine learning algorithms may seem daunting, it need not be so. By sticking to the priority of outcome transparency, you and your team will be able to follow some straightforward and simple guidelines for selecting sufficiently interpretable but optimally performing algorithmic techniques.

Before exploring these guidelines, it is necessary to provide you with some background information to help you better understand what facets of explanation are actually involved in technically interpretable AI. A good grasp of what is actually needed from such an explanation will enable you to effectively target the interpretability needs of your AI project.

**Facets of explanation in technically interpretable AI:** A good starting point for understanding how the technical dimension of explanation works in interpretable AI systems is to remember that these systems are largely mathematical models that carry out step-by-step computations in transforming sets of statistically interacting or independent inputs into sets of target outputs. Machine learning is, at bottom, just applied statistics and probability theory fortified with several other mathematical techniques. As such, it is subject to same methodologically rigorous requirements of logical validation as other mathematical sciences.

Such a demand for rigour informs the facet of **formal and logical explanation of AI systems** that is sometimes called the ***mathematical glass box***. This characterisation refers to the transparency of strictly formal explanation: No matter how complicated it is (even in the case of a deep neural net with a hundred million parameters), an algorithmic model is a closed system of effectively computable operations where rules and transformations are mechanically applied to inputs to determine outputs. In this restricted sense, all AI and machine learning models are fully intelligible and mathematically transparent if only ***formally and logically*** so.

This is an important characteristic of AI systems, because it makes it possible for supplemental and eminently interpretable computational approaches to model, approximate, and simplify even the most complex and high dimensional among them. In fact, such a possibility fuels some of the technical approaches to interpretable AI that will soon be explored.

This formal way of understanding the technical explanation of AI and machine learning systems, however, has immediate limitations. It can tell us that a model is mathematically intelligible because it operates according to a collection of fixed operations and parameters, but it cannot tell us much about how or why the components of the model transformed a specified group of inputs into their corresponding outputs. It cannot tell us anything about the *rationale behind the algorithmic generation of a given outcome*.

This second dimension of technical explanation has to do with the *semantic facet* of interpretable AI. A **semantic explanation** offers an interpretation of the functions of the individual parts of the algorithmic system in the generation of its output. Whereas formal and logical explanation presents an account of the stepwise application of the procedures and rules that comprise the formal framework of the algorithmic system, semantic explanation helps us to understand the meaning of those procedures and rules in terms of their purpose in the input-output mapping operation of the system, i.e. what role they play in determining the outcome of the model's computation.

The difficulties surrounding the interpretability of algorithmic decisions and behaviours arise in this semantic dimension of technical explanation. It is easiest to illustrate this by starting from the simplest case.

When a machine learning model is very basic, the task of following the rationale of how it transforms a given set of inputs into a given set of outputs can be relatively unproblematic. For instance, in the simple linear regression, $y = a + bx + \varepsilon$, with a single predictor variable $x$ and a response variable $y$, the predictive relationship of x to y is directly expressed in a regression coefficient $b$, representing the rate and direction at which $y$ is predicted to change as x changes. This hypothetical model is completely interpretable from the technical perspective for the following reasons:

- **Linearity:** Any change in the value of the predictor variable is directly reflected in a change in the value of the response variable at a constant rate $b$. The interpretable prediction yielded by the model can therefore be directly inferred. This linearity dimension of predictive models has been an essential feature of the automated decision-making systems in many heavily regulated and high-impact sectors, because the predictions yielded have high inferential clarity and strength.

- **Monotonicity:** When the value of the predictor changes in a given direction, the value of the response variable changes consistently either in the same or opposite direction. The interpretable prediction yielded by the model can thus be directly inferred. This monotonicity dimension is also a highly desirable interpretability condition of predictive models in many heavily regulated sectors, because it incorporates reasonable expectations about the consistent application of sector specific selection constraints into automated decision-making systems. So, for example, if the selection criteria to gain employment at an agency or firm includes taking an exam, a reasonable expectation of outcomes would be that if candidate A scored better than candidate B, then candidate B, all other things being equal, would not be selected for employment when A is not. A monotonic predictive model that uses the exam score as the predictor variable and application success as the response variable would, in effect, guarantee this expectation is met by disallowing situations where A scores better than B but B gets selected and A does not.

- **Non-Complexity:** The number of features (dimensionality) and feature interactions is low enough and the mapping function is simple enough to enable a clear 'global' understanding of the function of each part of the model in relation to its outcome.

While, all three of these desirable interpretability characteristics of the imagined model allow for direct and intuitive reasoning about the relation of the predictor and response variables, the model itself is clearly too minimal to capture the density of relationships and interactions between attributes in complex real-world situations where some degree of noisiness is unavoidable and the task of apprehending the subtleties of underlying data distributions is tricky.

In fact, one of the great strides forward that has been enabled by the contemporary convergence of expanding computing power and big data availability with more advanced machine learning models has been exactly this capacity to better capture and model the intricate and complicated dynamics of real-world situations. Still, this incorporation of the complexity of scale into the models themselves has also meant significant challenges to the semantic dimension of the technical explanation of AI systems.

As machine learning systems have come to possess both ever greater access to big data and increasing computing power, their designers have correspondingly been able both to enlarge the feature spaces (the number of input variables) of these systems and to turn to gradually more complex mapping functions. In many cases, this has meant vast improvements in the predictive and classificatory performance of more accurate and expressive models, but this has also meant the growing prevalence of non-linearity, non-monotonicity, and high-dimensional complexity in an expanding array of so-called 'black-box' models.

Once high-dimensional feature spaces and complex functions are introduced into machine learning systems, the effects of changes in any given input become so entangled with the values and interactions of other inputs that understanding how individual components are transformed into outputs becomes extremely difficult. The complex and unintuitive curves of the decision functions of many of these models preclude linear and monotonic relations between their inputs and outputs. Likewise, the high-dimensionality of their optimisation techniques—frequently involving millions of parameters and complex correlations—ranges well beyond the limits of human-scale cognition and understanding. To illustrate the increasing complexity involved in comprehending input-output mappings, here is a visual representation that depicts the difference of between a linear regression function and a deep neural network:

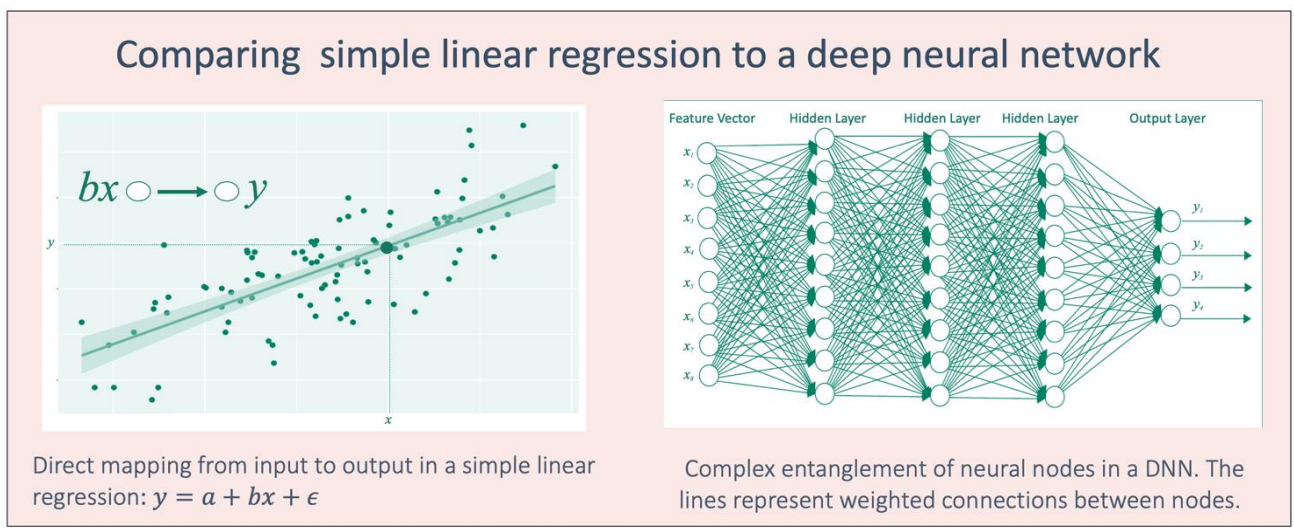

Comparing simple linear regression to a deep neural network

Direct mapping from input to output in a simple linear regression: $y = a + bx + \epsilon$

Complex entanglement of neural nodes in a DNN. The lines represent weighted connections between nodes.

These rising tides of computational complexity and algorithmic opacity consequently pose a key challenge for the responsible design and deployment of safe, fair, and ethical AI systems: how should the potential to advance the public interest through the implementation of high performing but increasingly uninterpretable machine learning models be weighed against the tangible risks posed by the lack of interpretability of such systems?

A careful answer to this question is, in fact, not so simple. While the trade-off between performance and interpretability may be real and important in *some domain-specific applications*, in others there exist increasingly sophisticated developments of standard interpretable techniques such as regression extensions, decision trees, and rule lists that may prove just as effective for use cases where the need for transparency is paramount. Furthermore, supplemental interpretability tools, which function to make 'black box' models more semantically and qualitatively explainable are rapidly advancing day by day.

These are all factors that you and your team should consider as you work together to decide on which models to use for your AI project. As a starting point for those considerations, let us now turn to some basic guidelines that may help you to steer that dialogue toward points of relevance and concern.

## Guidelines for designing and delivering a sufficiently interpretable AI system

You should use the table below to begin thinking about how to integrate interpretability into your AI project. While aspects of this topic can become extremely technical, it is important to make sure that dialogue about making your AI system interpretable remains multidisciplinary and inclusive. Moreover, it is crucial that key stakeholders be given adequate consideration when deciding upon the delivery mechanisms of your project. These should include policy or operational design leads, the technical personnel in charge of operating the trained models, the implementers of the models, and the decision subjects, who are affected by their outcomes.

Note that the first three guidelines focus on the big picture issues you will need to consider in order to incorporate interpretability needs into your project planning and workflow, whereas the last two guidelines shift focus to the user-centred requirements of designing and implementing a sufficiently interpretable AI system.

| Guidelines for designing and delivering a sufficiently interpretable AI system |
|---|
| **Guideline 1:** Look first to context, potential impact, and domain-specific need when determining the interpretability requirements of your project |
| There are several related factors that should be taken into account as you formulate your project's approach to interpretability: |
| 1. **Type of application:** Start by assessing both the kind of tool you are building and the environment in which it will apply. Clearly there is a big difference between a computer vision system that sorts handwritten employee feedback forms and one that sorts safety risks at a security checkpoint. Likewise, there is a big difference between a random forest model that triages applicants at a licencing agency and one that triages sick patients in an emergency department.<br><br>Understanding your AI system's purpose and context of application will give you a better idea of the stakes involved in its use and hence also a good starting point to think about the scope of its interpretability needs. For instance, low-stakes AI models that are |

not safety-critical, do not directly impact the lives of people, and do not process potentially sensitive social and demographic data will likely have a lower need for extensive resources to be dedicated to a comprehensive interpretability platform.

2. **Domain specificity:** By acquiring solid domain knowledge of the environment in which your AI system will operate, you will gain better insight into any potential sector-specific standards of explanation or benchmarks of justification which should inform your approach to interpretability. Through such knowledge, you may also obtain useful information about organisational and public expectations regarding the scope, content, and depth of explanations that have been previously offered in relevant use cases.

3. **Existing technology:** If one of the purposes of your AI project is to replace an existing algorithmic technology that may not offer the same sort of expressive power or performance level as the more advanced machine learning techniques that you are planning to deploy, you should carry out an assessment of the performance and interpretability levels of the existing technology. Acquiring this knowledge will provide you with an important reference point when you are considering possible trade-offs between performance and interpretability that may occur in your own prospective system. It will also allow you to weigh the costs and benefits of building a more complex system with higher interpretability-support needs in comparison to the costs and benefits of using a simpler model.

Guideline 2: Draw on standard interpretable techniques when possible

In order to actively integrate the aim of sufficient interpretability into your AI project, your team should approach the model selection and development process with the goal of finding the right fit between **(1) domain-specific risks and needs, (2) available data resources and domain knowledge, and (3) task appropriate machine learning techniques**. Effectively assimilating these three aspects of your use case requires open-mindedness and practicality.

Often times, it may be the case that high-impact, safety-critical, or other potentially sensitive environments heighten demands for the thoroughgoing accountability and transparency of AI projects. In some of these instances, such demands may make choosing standard but sophisticated non-opaque techniques an overriding priority. These techniques may include **decisions trees, linear regression and its extensions like generalised additive models, decision/rule lists, case-based reasoning, or logistic regression**. In many cases, reaching for the 'black box' model first may not be appropriate and may even lead to inefficiencies in project development, because more interpretable models, which perform very well but do not require supplemental tools and techniques for facilitating interpretable outcomes, are also available.

Again, solid domain knowledge and context awareness are key components here. In use cases where data resources lend to well-structured, meaningful representations and domain expertise can be incorporated into model architectures, interpretable techniques may often be more desirable than opaque ones. Careful data pre-processing and iterative model development can, in these cases, hone the accuracy of such interpretable systems in ways that may make the advantages gained by the combination of their performance and transparency outweigh the benefits of more semantically intransparent approaches.

In other use cases, however, data processing needs may disqualify the deployment of these sorts of straightforward interpretable systems. For instance, when AI applications are sought for classifying images, recognising speech, or detecting anomalies in video footage, the most effective machine learning approaches will likely be opaque. The feature spaces of these kinds of AI systems grow exponentially to hundreds of thousands or even millions of dimensions. At this scale of complexity, conventional methods of interpretation no longer apply. Indeed, it is the unavoidability of hitting such an **interpretability wall** for certain important applications of supervised, unsupervised, and reinforcement learning that has given rise to an entire subfield of machine learning research which focuses on providing technical tools to facilitate interpretable and explainable AI.

When the use of 'black box' models best fits the purpose of your AI project, you should proceed diligently and follow the procedures recommended in Guideline 3. For clarity, let us define a 'black box' model as **any AI system whose innerworkings and rationale are opaque or inaccessible to human understanding**. These systems may include **neural networks** (including recurrent, convolutional, and deep neural nets), **ensemble methods** (an algorithmic technique such as the random forest method that strengthens an overall prediction by combining and aggregating the results of several or many different base models), and **support vector machines** (a classifier that uses a special type of mapping function to build a divider between two sets of features in a high dimensional feature space).

Guideline 3: When considering the use of 'black box' AI systems, you should:

1. Thoroughly weigh up impacts and risks;

2. Consider the options available for supplemental interpretability tools that will ensure a level of semantic explanation which is both *domain appropriate* and *consistent with the design and implementation of safe, fair, and ethical AI*;

3. Formulate an interpretability action plan, so that you and your team can put adequate forethought into how explanations of the outcomes of your system's decisions, behaviours, or problem-solving tasks can be optimally provided to users, decision subjects, and other affected parties.

It may be helpful to explore each of these three suggested steps of assessing the viability of the responsible design and implementation of a 'black box' model in greater detail.

**(1) Thoroughly weigh up impacts and risks:** Your first step in evaluating the feasibility of using a complex AI system should be to focus on issues of ethics and safety. As a general policy, you and your team should utilise 'black box' models only:

- where their potential impacts and risks have been thoroughly considered in advance, and you and your team have determined that your use case and domain specific needs support the responsible design and implementations of these systems;

- where supplemental interpretability tools provide your system with a domain appropriate level of semantic explainability that is reasonably sufficient to mitigate its potential risks and that is therefore consistent with the design and implementation of safe, fair, and ethical AI.

**(2) Consider the options available for supplemental interpretability tools:** Next, you and your team should assess whether there are technical methods of explanation-support that ***both*** satisfy the specific interpretability needs of your use case as determined by the deliberations suggested in Guideline 1 ***and*** are appropriate for the algorithmic approach you intend to use. You should consult closely with your technical team at this stage of model selection. The exploratory processes of trial-and-error, which often guide this discovery phase in the innovation lifecycle, should be informed and constrained by a solid working knowledge of the technical art of the possible in the domain of available and useable interpretability approaches.

The task of lining up the model selection process with the demands of interpretable AI requires a few conceptual tools that will enable thoughtful evaluation of whether proposed supplemental interpretability approaches sufficiently meet your project's explanatory needs. First and most importantly, you should be prepared to ask the right questions when evaluating any given interpretability approach. This involves establishing with as much clarity as possible **how the explanatory results of that approach can contribute to the user's ability to offer solid, coherent, and reasonable accounts of the rationale behind any given algorithmically generated output**. Relevant questions to ask that can serve this end are:

- What sort of explanatory resources will the interpretability tool provide users and implementers in order (1) to enable them to exercise better-informed evidence-based judgments and (2) to assist them in offering plausible, sound, and reasonable accounts of the logic behind algorithmically generated output to affected individuals and concerned parties?

- Will the explanatory resources that the interpretability tool offers be useful for providing affected stakeholders with a sufficient understanding of a given outcome?

- How, if at all, might the explanatory resources offered by the tool be misleading or confusing?

You and your team should take these questions as a starting point for evaluating prospective interpretability tools. These tools should be assessed in terms of their capacities to render the reasoning behind the decisions and behaviours of the uninterpretable 'black box' systems sufficiently intelligible to users and affected stakeholders given use case and domain specific interpretability needs.

Keeping this in mind, there are two technical dimensions of supplemental interpretability approaches that should be systematically incorporated into evaluation processes at this stage of the innovation workflow.

The first involves the possible **explanatory strategies** you choose to pursue over the course of the design and implementation lifecycle. Such strategies will largely determine the paths to understanding you will be able to provide for its users and decision subjects. They will largely define *how you explain your model and its outcomes* and hence *what kinds of explanation you are able offer*.

The second involves the **coverage and scope** of the actual explanations themselves. The choices you make about explanatory coverage will determine the extent to which the kinds of explanations you are planning to pursue will address *single instances* of the model's outputs or range more broadly to cover the *underlying rationale of its behaviour in general and across instances*. Choices you make about explanatory coverage will largely govern the extent to which your AI system is locally and/or globally interpretable.

The very broad-brushed overview of these two dimensions that follows is just meant to orient you to some of the basic concepts in an expanding field of research, so that you are more prepared for working with your technical team to think through the strengths and weaknesses of various approaches. Note, additionally, that this is a rapidly developing area. Relevant members of your team should keep abreast of the latest developments in the field of interpretable AI or XAI (Explainable AI):

**Two technical dimensions of supplemental interpretability approaches:**

1. **Determining explanatory strategies:** To achieve the goal of securing a sufficiently interpretable AI system, you and your team will need to get clear on **how to explain** your model and its outcomes. The explanatory strategies you decide to pursue will shape the paths to understanding you are able to provide for the users of your model and for its decision subjects.

   There are four such explanatory strategies to which you should pay special attention:

   a) *Internal explanation:* Pursuing the internal explanation of an opaque model involves making intelligible how the components and relationships within it function. There are two ways that such a goal of internal explanation can be interpreted. On the one hand, it can be seen as an endeavour to explain the operation of the model by considering it globally *as a comprehensible whole*. Here, the aspiration is to 'pry open the black box' by building an explanatory model that enables a full grasp of the opaque system's internal contents. The strengths and weaknesses of such an approach will be discussed in the next section on global interpretability.

   On the other hand, the search for internal explanation can indicate the pursuit a kind of *engineering insight*. In this sense, internal explanation can be seen as attempting to shed descriptive and inferential light on the parts and operation of the system as a whole in order to try to make it work better. Acquiring this sort of internal understanding of the more general relationships that the working parts of a trained model have with patterns of its responses can allow researchers to advance step-by-step in gaining a better data scientific grasp on why it does what it does and how to improve it. Similarly, this type of internal explanation can be seen as attempting to shed light on an opaque model's operation by breaking it down into more understandable, analysable, and digestible parts (for instance, in the case of a DNN: into interpretable characteristics of its vectors, features, layers, parameters, etc.).

From a practical point of view, this kind of aspiration to **engineering insight** in the ends of data scientific advancement should inform the goals of your technical team throughout the model selection and design workflow. Numerous methods exist to help provide informative representations of the innerworkings of various 'black box' systems. Gaining a clearer descriptive understanding of the internal composition of your system will contribute greatly to your project's ability to achieve a higher degree of outcome transparency and to its capacity to foster best practices in the pursuit of responsible data science in general.

b) *External or post-hoc explanation:* External or post-hoc explanation attempts to capture essential attributes of the observable behaviour of a 'black box' system by subjecting it to a number of different techniques that reverse engineer explanatory insight. Some post-hoc approaches test the sensitivity of the outputs of an opaque model to perturbations in its inputs; others allow for the interactive probing of its behavioural characteristics; others, still, build proxy-based models that utilise simplified interpretable techniques to gain a better understanding of particular instances of its predictions and classifications.

This external or post-hoc approach has, at present, established itself in machine learning research as a go-to explanatory strategy and for good reason. It allows data scientists to pose mathematical questions to their opaque systems by testing them and by building supplemental models which enable greater insight through the inferences drawn from their experimental interventions. Such a post-hoc approach allows them, moreover, to seek out evidence for the reasoning behind a given opaque model's prediction or classification by utilising maximally interpretable techniques like linear regression, decision trees, rule lists, or case-based reasoning. Several examples of post-hoc explanation will be explored below in the section on local interpretability.

Take note initially though that, as some critics have rightly pointed out, because they are approximations or simplified supplemental models of the more complex originals, many post-hoc explanations can fail to accurately represent certain areas of the opaque model's feature space. This deterioration of accuracy in parts of the original model's domain can frequently produce misleading and uncertain results in the post-hoc explanations of concern.

c) *Supplemental explanatory infrastructure*: A different kind of explanatory strategy involves actually incorporating secondary explanatory facilities into the system you are building. For instance, an image recognition system could have a primary component, like a convolutional neural net, that extracts features from its inputs and classifies them while a secondary component, like a built-in recurrent neural net with an 'attention-directing' mechanism, translates the extracted features into a natural language representation that produces a sentence-long explanation of the result to the user. In other words, a system like this is designed to provide simple explanations of its own data processing results.

Research into integrating 'attention-based' interfaces like this in AI systems is continuing to advance toward making their implementations more sensitive to user needs, more explanation-forward, and more human-understandable. For instance, multimodal methods of combining visualisation tools and textual interface are being developed that may make the provision of explanations more interpretable for both implementers and decision subjects. Furthermore, the incorporation of domain knowledge and logic-based or convention-based structures into the architectures of complex models are increasingly allowing for better and more user-friendly representations and prototypes to be built into them. This is gradually enabling more sophisticated explanatory infrastructures to be integrated into opaque systems and makes it essential to think about building explanation-by-design into your AI projects.

d) **Counterfactual explanation**: While counterfactual explanation is a kind of post-hoc approach, it deserves special attention insofar as it moves beyond other post-hoc explanations to provide affected stakeholders with clear and precise options for actionable recourse and practical remedy.

Counterfactual explanations are contrastive explanations: They offer succinct computational reckonings of how specific factors that influenced an algorithmic decision can be changed so that better alternatives can be realised by the subject of that decision. Incorporating counterfactual explanations into your AI system at its point of delivery would allow stakeholders to see what input variables of the model can be modified, so that the outcome could be altered to their benefit. Additionally, from a responsible design perspective, incorporating counterfactual explanation into the development and testing phases of your system would allow your team to build a model that incorporates *actionable variables*, i.e. input variables that will afford decision subjects with concise options for making practical changes that would improve their chances of obtaining the desired outcome. **Counterfactual explanatory strategies can be used as way to incorporate reasonableness and the encouragement of agency into the design and implementation of your AI project**.

All that said, it is important to recognise that, while counterfactual explanation does offer an innovative way to contrastively explore how feature importance may influence an outcome, it is not a complete solution to the problem of AI interpretability. In certain cases, for instance, the sheer number of potentially significant features that could be at play in counterfactual explanations of a given result can make a clear and direct explanation difficult to obtain and selected sets of explanations seem potentially arbitrary. Moreover, there are as yet limitations on the types of datasets and functions to which these kinds of explanations are applicable. Finally, because this kind of explanation concedes the opacity of the algorithmic model outright, it is less able to address concerns about potentially harmful feature interactions and multivariate relationships that may be buried deep within the model's architecture.

Here is an at-a-glance view of a typology of these explanatory strategies:

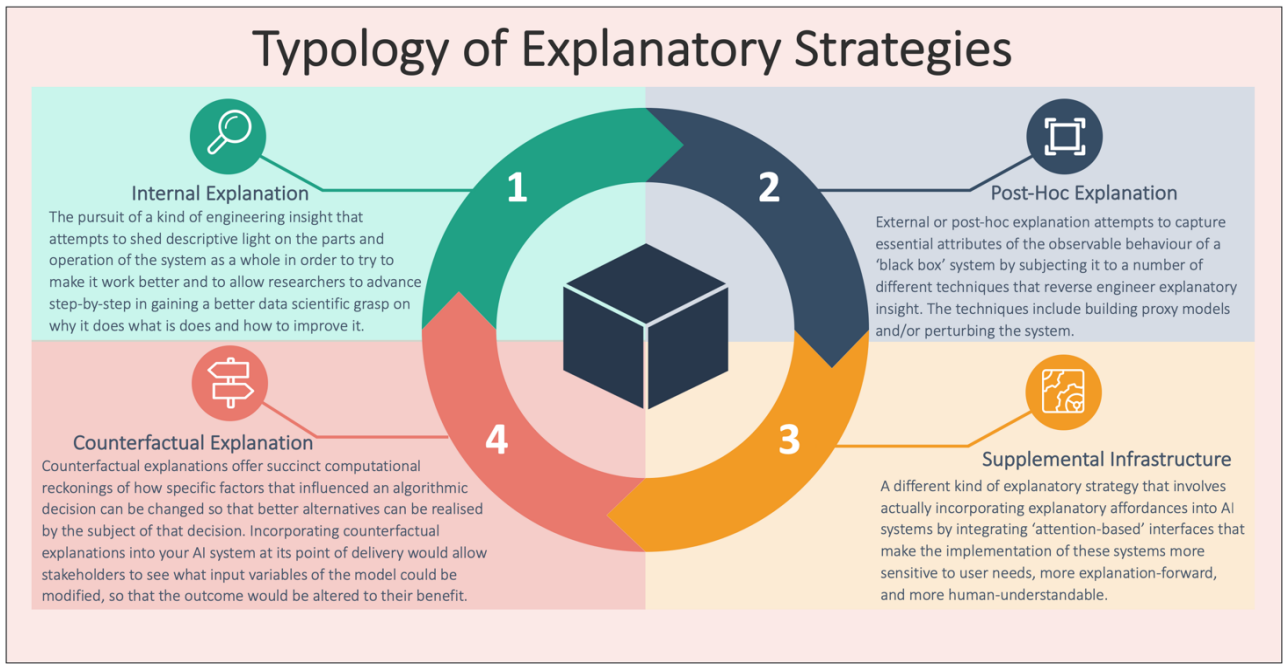

2. **Coverage and Scope:** The main questions you will need to broach in the dimension of the coverage and scope of your supplemental interpretability approach are: To what extent does our interpretability approach cover the explanation of *singe predictions or classifications* of the model and to what extent does it cover the explanation of the *innerworkings and rationale of the model as a whole and across predictions*? To what extent does it cover both?

   This distinction between single instance and total model explanation is often characterised as the difference between **local interpretability** and the **global interpretability**. Both types of explanation offer potentially helpful support for the provision of significant information about the rationale behind an algorithmic decision or behaviour, but both, in their own ways, also face difficulties.

   **Local Interpretability:** A local semantic explanation aims to enable the interpretability of **individual cases**. The general idea behind attempts to explain a 'black box' system in terms of specific instances is that, regardless of how complex the architecture or decision function of that system may be, it is possible to gain interpretive insight into its innerworkings by focusing on single data points or neighbourhoods in its feature space. In other words, even if the high dimensionality and curviness of a model makes it opaque *as a whole*, there is an expectation that insight-generating interpretable methods can be applied *locally* to smaller sections of the model, where changes in isolated or grouped variables are more manageable and understandable.

This general explanatory perspective has yielded several different interpretive strategies that have been successfully applied in significant areas of 'black box' machine learning. One family of such strategies has zeroed in on neural networks (DNNs, in particular) by identifying what features of an input vector's data points make it representative of the target concept that a given model is trying to classify. So, for example, if we have a digital image of a dog that is converted into a vector of pixel values and then processed it through a dog-classifying deep neural net, this interpretive approach will endeavour to tell us why the system yielded a 'dog-positive' output by isolating the slices of this set of data points that are most relevant to its successful classification by the model.

This can be accomplished in several related ways. What is called **sensitivity analysis** identifies the most relevant features of an input vector by calculating local gradients to determine how a data point has to be moved to change the output label. Here, an output's sensitivity to such changes in input values identifies the most relevant features. Another method to identify feature relevance that is downstream from sensitivity analysis is called **salience mapping**, where a strategy of moving backward through the layers of a neural net graph allows for the mapping of patterns of high activation in the nodes and ultimately generates interpretable groupings of salient input variables that can be visually represented in a heat or pixel attribution map.

A second local interpretive strategy also seeks to explain feature importance in a single prediction or classification by perturbing input variables. However, instead of using these nudges in the feature space to highlight areas of saliency, it uses them to prod the opaque model in the area around the relevant prediction, so that a supplemental interpretable model can be constructed which establishes the relative importance of features in the black box model's output.

The most well-known example of this strategy is called **LIME (Local Interpretable Model-Agnostic Explanation)**. LIME works by fitting an interpretable model to a specific prediction or classification produced by the opaque system of concern. It does this by sampling data points at random around the target prediction or classification and then using them to build a local approximation of the decision boundary that can account for the features which figure prominently in the specific prediction or classification under scrutiny.

The way this works is relatively uncomplicated: LIME generates a simple linear regression model by weighting the values of the data points, which were produced by randomly perturbing the opaque model, according to their proximity to the original prediction or classification. The closest of these values to the instance being explained are weighted the heaviest, so that the supplemental model can produce an explanation of feature importance that is **locally faithful** to that instance.  Note that the type of model that LIME uses most prominently is a sparse linear regression function for reasons of semantic transparency that were discussed above. Other interpretable models such as decision trees can likewise be employed.

While LIME does indeed appear to be a step in the right direction for the future of interpretable AI, a host of issues that present challenges to the approach remains unresolved. For instance, the crucial aspect of how to properly define the proximity measure for the 'neighbourhood' or 'local region' where the explanation applies remains unclear, and small changes in the scale of the chosen measure can lead to greatly diverging explanations. Likewise, the explanation produced by the supplemental linear model can quickly become unreliable even with small and virtually unnoticeable perturbations of the system it is attempting to approximate. This challenges the basic assumption that that there is always some simplified linear model that successfully approximates the underlying model reasonably well near any given data point.

LIME's creators have largely acknowledged these shortcomings and have recently offered a new explanatory approach that they call '**anchors**'. These 'high precision rules' incorporate into their formal structures 'reasonable patterns' that are operating within the underlying model (such as the implicit linguistic conventions that are at work in a sentiment prediction model), so that they can establish suitable and faithful boundaries of their explanatory coverage of its predictions or classifications.

A related and equally significant local interpretive strategy is called **SHAP (Shapley Additive exPlanations).** SHAP uses concepts from game theory to define a 'Shapley value' for a feature of concern that provides a measurement of its influence on the underlying model's prediction. Broadly, this value is calculated for a feature by averaging its marginal contribution to *every possible prediction* for the instance under consideration.

This might seem impossible, but the strategy is straightforward. SHAP calculates the marginal contribution of the relevant feature for all possible combinations of inputs in the feature space of the instance. So, if the opaque model that it is explaining has 15 features, SHAP would calculate the marginal contribution of the feature under consideration 32,768 times (i.e. one calculation for each combination of all possible combinations of features: $2^{15}$, or $2^k$ when $k = 15$).

This method then allows SHAP to estimate the Shapley values for all input features in the set to produce the complete distribution of the prediction for the instance. In our example, this would entail 491,520 calculations. While such a procedure is computationally burdensome and becomes intractable beyond a certain threshold, this means that *locally*, that is, for the calculation of the specific instance, SHAP can axiomatically guarantee the consistency and accuracy of its reckoning of the marginal effect of the feature. (Note that the SHAP platform does offer methods of approximation to avoid this excessive computational expense.)

Despite this calculational robustness, SHAP also faces some of the same kinds of difficulties that LIME does. The way SHAP calculates marginal contributions is by constructing two instances: the first instance includes the feature being measured while the second leaves it out. After calculating the prediction for each of these instances by plugging their values into the underlying model, the result of the second is subtracted from that of the first to determine the marginal contribution of the feature. This procedure is then repeated for all possible combinations of features so that the weighted average of all of the marginal contributions of the feature of concern can be computed.

The contestable part of this process comes with how SHAP defines the **absence** of variables under consideration. To leave out a feature—whether it's the one being directly measured or one of the others not included in the combination under consideration—SHAP replaces it with a *stand-in feature value* drawn from a selected donor sample (that is itself drawn from the existing dataset). This method of sampling values assumes feature independence (i.e. that values sampled are not correlated in ways that might significantly affect the output for a particular calculation). As a consequence, the interaction effects engendered by and between stand-in variables are necessarily unaccounted for when conditional contributions are approximated. The result is the introduction of uncertainty into the explanation that is produced because the complexity of multivariate interactions in the underlying model may not be sufficiently captured by the simplicity of this supplemental interpretability technique. This drawback in sampling (as well as a certain degree of arbitrariness in domain definition) can cause SHAP to become unreliable even with minimal perturbations of the model it is approximating.

Despite these limitations in the existing tools of local interpretability, it is important that you think 'local-first' when considering the issue of the coverage and scope of the explanatory approaches you plan to incorporate into your project. Being able to provide explanations of specific predictions and classifications is of paramount importance both to securing optimal outcome transparency and also to ensuring that your AI system will be implemented responsibly and reasonably.

**Global interpretability:** The motivation behind the creation of local interpretability tools like LIME or SHAP (as well as many others not mentioned here) has derived, at least in part, from a need to find a way of avoiding the kind of difficult *double bind* faced by the alternative approach to the coverage and scope of interpretable AI: global interpretability.

On the prevailing view, providing a global explanation of a 'black box' model entails offering an alternative interpretable model that captures the innerworkings and logic of a 'black box' model *in sum* and across predictions or classifications. The difficulty faced by global interpretability arises in the seemingly unavoidable trade-off between the need for the global explanatory model to be sufficiently simple so that it is understandable by humans and the need for that model to be sufficiently complex so that it can capture the intricacies of how the mapping function of a 'black box' model works as a whole.

While this is clearly a real problem that appears to be theoretically inevitable, it is important to keep in mind that, *from a practical standpoint*, a serviceable notion of global interpretability need not be limited to such a conceptual puzzle. There are at least two less ambitious but more constructive ways to view global interpretability as a potentially meaningful contributor to the responsible design and implementation of interpretable AI.

First, many useful attempts have already been made at building explanatory models that employ interpretable methods (like decision trees, rule lists, and case-based classification) to globally approximate neural nets, tree ensembles, and support vector machines. These results have enabled a deeper understanding of the way human interpretable logics and conventions (like if-then rules and representationally generated prototypes) can be measured against or mapped onto high dimensional computational structures and even allow for some degree of targeted comprehensibility of the logic of their parts.

This capacity to 'peek into the black box' is of great practical importance in domains where trust, user-confidence, and public acceptance are critical for the realisation optimal outcomes. Moreover, this ability to move back and forth between interpretable architectures and high-dimensional processing structures can enable knowledge discovery as well as insights into the kinds of dataset-level and population-level patterns, which are crucial for well-informed macroscale decision-making in areas ranging from public health and economics to the science of climate change.

Being able to uncover global effects and relationships between complex model behaviour and data distributions at the demographic and ecological level may prove vital for establishing valuable and practically useful knowledge about unobservable but significant biophysical and social configurations. Hence, although these models have not solved the understandability-complexity puzzle as such, they have opened up new pathways for innovative thinking in the applied data sciences that may be of immense public benefit in the future.

Secondly, as mentioned above, under the auspices of the aspiration to **engineering insight**, a ***descriptive and analytical kind of global interpretability*** can be seen as a driving force of data scientific advancement. When seen through a practitioner-centred lens, this sort of global interpretability allows data scientists to take a wide-angled and discovery-oriented view of a 'black box' model's relationship to patterns that arise across the range of its predictions. Figuring out how an opaque system works and how to make it work better by more fully understanding these patterns is a continuous priority of good research. So too is understanding the relevance of features and of their complex interactions through dataset level measurement and analysis. These dimensions of incorporating the explanatory aspirations of global interpretability into best practices of research and innovation should be encouraged in your AI project.

(3) **Formulate an interpretability action plan:** The final step you will need to take to ensure a responsible approach to using 'black box' models is to formulate an interpretability action plan so that you and your team can put adequate forethought into how explanations of the outcomes of your system's decisions, behaviours, or problem-solving tasks can be optimally provided to users, decision subjects, and other affected parties.

This action plan should include the following:

- A **clear articulation of the explanatory strategies** your team intends to use and a detailed plan that indicates the stages in the project workflow when the design and development of these strategies will need to take place.

- A succinct formulation of your **explanation delivery strategy**, which addresses the special provisions for clear, simple, and user-centred explication that are called for when supplemental interpretability tools for 'black box' models are utilised. See more about delivery and implementation in Guideline 5.

- A **detailed timeframe for evaluating your team's progress** in executing its interpretability action plan and a **role responsibility list**, which maps in detail the various task-specific responsibilities that will need to be fulfilled to execute the plan.

## Guideline 4: Think about interpretability in terms of the capacities of human understanding

When you begin to deliberate about the specific scope and content of your interpretability platform, it is important to reflect on what it is that you are exactly aiming to do in making your model sufficiently interpretable. A good initial step to take in this process is to think about what makes even the simplest explanations **clear and understandable**. In other words, you should begin by thinking about interpretability in terms of the capacities and limitations of human cognition.

From this perspective, it becomes apparent that even the most straightforward model like a linear regression function or a decision tree can become uninterpretable when its dimensionality presses beyond the cognitive limits of a thinking human. Recall our example of the simple linear regression: $y = a + bx + \epsilon$. In this instance, only one feature $x$ relates to the response variable $y$, so understanding the predictive relationship is easy. The model is parsimonious.

However, if we started to add more features as covariates, even though the model would remain linear and hence intuitively predictable, being able to understand the relationship between the response variable and all the predictors and their coefficients (feature weights) would quickly become difficult. So, say we added ten thousand features and trained the model: $y = a + b_0 x_0 + b_1 x_1 + \cdots + b_{10000} x_{10000} + \epsilon$. Understanding *how* this model's prediction comes about—what role each of the individual parts play in producing the prediction—would become difficult because of a certain cognitive limit in the quantity of entities that human thinking can handle at any given time. This model would lose a significant degree of interpretability.

Seeing interpretability as a continuum of comprehensibility that is dependent on the capacities and limits of the individual human interpreter should key you in to what is needed in order to deliver an interpretable AI system. Such limits to consider should include not only cognitive boundaries but also varying levels of access to relevant vocabularies of explanation; an explanation about the results of a trained model that uses a support vector machine to divide a 26-dimensional feature space with a planar separator, for instance, may be easy to understand for a technical operator or auditor but entirely inaccessible to a non-specialist. Offering good explanations should take expertise level into account. **Your interpretability platform should be cognitively equitable**.

## Securing responsible delivery through human-centred implementation protocols and practices

The demand for sensitivity to human factors should inform your approach to devising delivery and implementation processes from start to finish. To provide clear and effective explanations about the content and rationale of algorithmic outputs, you will have to begin by building *from the human ground up*. You will have to pay close attention to the circumstances, needs, competences, and capacities of the people whom your AI project aims to assist and serve.

This means that *context will be critical*. By understanding your use case well and by drawing upon solid domain knowledge, you will be better able to **define roles and relationships**. You will better be able to **train the users and implementers of your system**. And, you will be better able **to establish an effectual implementation platform, to clarify content, and to facilitate understanding of outcomes** for users and affected stakeholders alike. Here is a diagram of what securing human-centred implementation protocols and practices might look like:

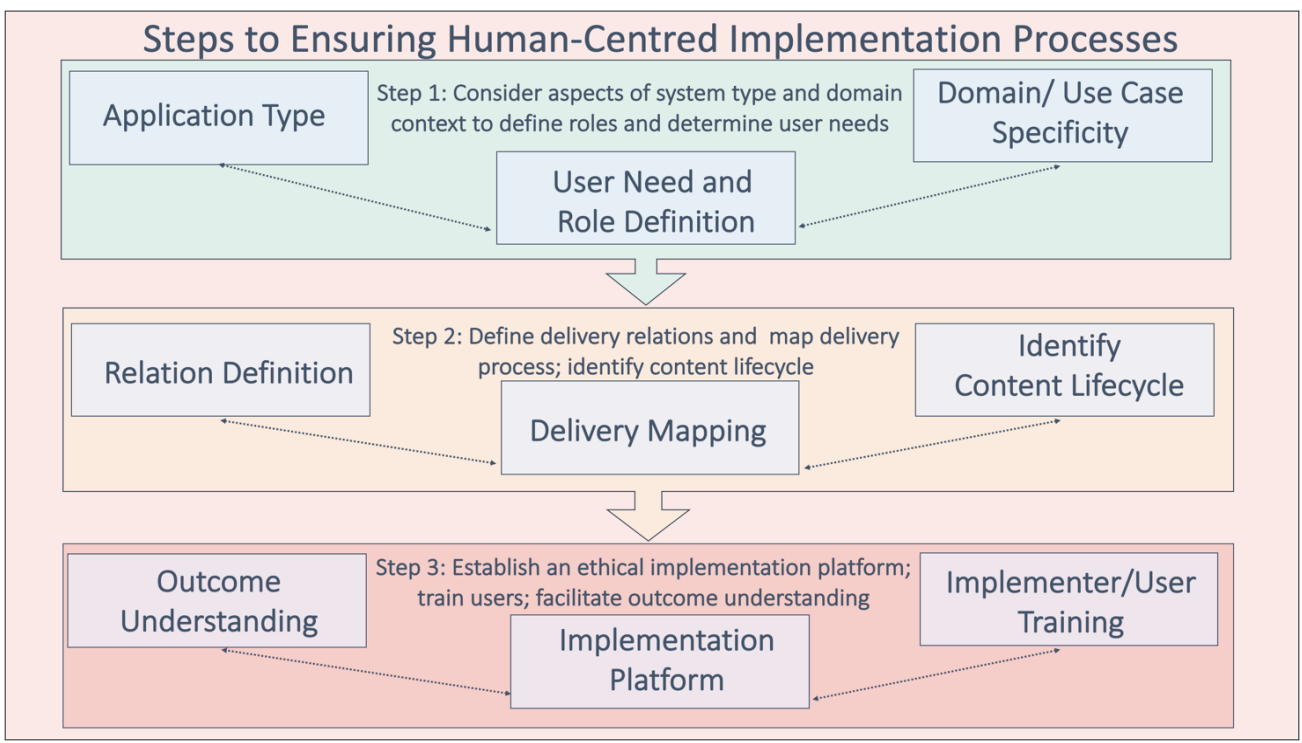

Let us consider these steps in turn by building a checklist of essential actions that should be taken to help ensure the human-centred implementation of your AI project. Because the specifics of your approach will depend so heavily on the context and potential impacts of your project, we'll assume a generic case and construct the checklist around a hypothetical algorithmic decision-making system that will be used for predictive risk assessment.

**Step 1**: Consider aspects of application type and domain context to define roles and determine user needs

☐ (1) Assess which members of the communities you are serving will be most affected by the implementation of your AI system. Who are the most vulnerable among them? How will their socioeconomic, cultural, and education backgrounds affect their capacities to interpret and understand the explanations you intend to provide? How can you fine-tune your explanatory strategy to accommodate their requirements and provide them with clear and non-technical details about the rationale behind the algorithmically supported result?

When thinking about providing explanations to affected stakeholders, you should start with the needs of the most disadvantaged first. Only in this way, will you be able to establish an acceptable baseline for the equitable delivery of interpretable AI.

☐ (2) After reviewing Guideline 1 above, make a list of and define all the roles that will potentially be involved at the delivery stage of your AI project. As you go through each role, specify levels of technical expertise and domain knowledge as well as possible goals and objectives for each role. For instance, in our predictive risk assessment case:

- o **Decision Subject (DS)-**
  - ▪ **Role:** Subject of the predictive analytics.
  - ▪ **Possible Goals and Objectives:** To receive a fair, unbiased, and reasonable determination, which makes sense; to discover which factors might be changed to receive a different outcome.
  - ▪ **Technical and Domain Knowledge:** Most likely low to average technical expertise and average domain knowledge.
- o **Advocate for the DS-**
  - ▪ **Role:** Support for the DS (for example, legal counsel or care worker) and concerned party to the automated decision.
  - ▪ **Possible Goals and Objectives:** To make sure the best interests of the DS are safeguarded throughout the process; to help make clear to the DS what is going on and how and why decisions are being made.
  - ▪ **Technical and Domain Knowledge:** Most likely average technical expertise and high level of domain knowledge.
- o **Implementer-**
  - ▪ **Role:** User of the AI system as decision support.
  - ▪ **Possible Goals and Objectives:** To make an objective and fair decision that is sufficiently responsive to the particular circumstances of the DS and that is anchored in solid reasoning and evidence-based judgement.
  - ▪ **Technical and Domain Knowledge:** Most likely average technical expertise and high level of domain knowledge.
- o **System Operator/Technician-**
  - ▪ **Role:** Provider of support and maintenance for the AI system and its use.

- **Possible Goals and Objectives:** To make sure the machine learning system is performing well and running in accordance with its intended design; to handle the technical dimension of information processing for the DS's particular case; to answer technical questions about the system and its results as they arise.
- **Technical and Domain Knowledge:** Most likely high level of technical expertise and average domain knowledge.

- **Delivery Manager-**
  - **Role:** Member of the implementation team who oversees its operation and responds to problems as they arise.
  - **Possible Goals and Objectives:** To ensure that the quality of the automation-supported assessment process is high and that the needs of the decision subject are being served as intended by the project; to oversee the overall quality of the relationships within the implementation team and between the members of that team and the communities they serve.
  - **Technical and Domain Knowledge:** Most likely average technical expertise and good to high level domain knowledge

**Step 2**: Define delivery relations and map delivery processes

- (1) Assess the possible relationships between the defined roles that will have significant bearing on your project's implementation and formulate a descriptive account of this relationship with an eye to the part it will play in the delivery process. For the predictive risk assessment example:

  - **Decision Subject/Advocate to Implementer**: This is the primary relationship of the implementation process. It should be information-driven and dialogue-driven with the implementer's exercise of unbiased judgment and the DS's comprehension of the outcome treated as the highest priorities. Implementers should be prepared to answer questions and to offer evidence-based clarifications and justifications for their determinations. The achievement of well-informed mutual understanding is a central aim.
  - **Implementer to System Operator**: This is the most critical operational relationship within the implementation team. Communication levels should be kept high from case to case, and the shared goal of the two parties should be to optimise the quality of the decisions by optimising the use of the algorithmic decision-support system in ways that are accessible both to the user and to the DS. The conversations between implementers and system operators should be problem-driven and should avoid, as much as possible, focus on the specialised vocabularies of each party's domain of expertise.
  - **Delivery Manager to Operator to Implementer**: The quality of this cross-disciplinary relationship within the implementation team will have direct bearing on the overall quality of the delivery of the algorithmically supported decisions. Safeguarding the latter will require that open and easily accessible lines of communication be maintained between delivery managers, operators, and implementers, so that unforeseen implementation problems can be tackled from multiple angles and in ways that anticipate and stem future difficulties. Additionally, different use cases may present different explanatory challenges that are best addressed by multidisciplinary team input. Good communications within the implementation team will be essential to enable that such challenges are addressed in a timely and efficient manner.

☐ (2) Start building a map of the delivery process. This should involve incorporating your understanding of the needs, roles, and relationships of relevant actors involved in the implementation of your AI system into the wider objective of providing clear, informative, and understandable explanations of algorithmically supported decisions.

It is vital to recognise, at this implementation-planning stage of your project, that the principal goal of the delivery process is two-fold: *to translate statistically expressed results into humanly significant reasons and to translate algorithmic outputs into socially meaningful outcomes*.

These overlapping objectives should have a direct bearing on the way you build a map for your project's delivery process, because they organise the duties of implementation into two task-specific components:

1. A **technical component**, which involves determining the most effective way to convey and communicate to users and decision subjects the statistical results of your model's information processing so that the factors that figured into the logic and rationale of those results can be translated into understandable reasons that can be subjected to rational evaluation and critical assessment; and

2. A **social component**, which involves clarifying the socially meaningful content of the outcome of a given algorithmically assisted decision by translating that model's technical machinery—its input and output variables, parameters, and functional rationale—*back* into the everyday language of the humanly relevant categories and relationships that informed the formulation of its purpose, objective, and intended elements of design in the first place. Only through this re-translation will the effects of that model's output on the real human life it impacts be understandable in terms of the specific social and individual context of that life and be conveyable as such.

These two components of the delivery process will be fleshed out in turn.

**Technical component of responsible implementation:** As a general rule, we use the results of statistical analysis to guide our actions, because, when done properly, this kind of analysis offers a solid basis of empirically derived evidence that helps us to exercise sound and well-supported judgment about the matters it informs.

Having a good understanding of the factors that are at work in producing the result of a particular statistical analysis (such as in an algorithmic decision-support system) means that we are able to grasp these factors (for instance, input features that weigh heavily in determining a given algorithmically generated decision) as reasons that may warrant the rational acceptability of that result. After all, seen from the perspective of the interpretability of such an analysis, these factors are, in fact, nothing other than *reasons that are operating to support its conclusions*.

Clearly understood, these factors that lie behind the logic of the result or decision are not 'causes' of it. Rather, they form the evidentiary basis of its rational soundness and of the goodness of the inferences that support it. Whether or not we ultimately agree with the decision or the result of the analysis, the reasons that work together to comprise its conclusions make **claims to validity** and can *as such* be called before a tribunal of **rational criticism**. These reasons, in other words, must bear the burden of continuous assessment, evaluation, and contestation.

This is an element especially crucial for the responsible implementation of AI systems: **Because they serve surrogate cognitive functions in society, their decisions and results are in no way immune from these demands for rational justification and thus must be delivered to be optimally responsive to such demands.**

The results of algorithmic decision support systems, in this sense, serve as stand-ins for acts of speech and representation and therefore bear the justificatory burdens of those cognitive functions. They must establish the validity of their conclusions and operate under the constraint of being surrogates of the dialogical goal to convince through good reasons.

This charge to be responsive to the demands of rational justification should be essential to the way you map out your delivery strategy. **When you devise how best to relay and explain the statistical results of your AI systems, you need to start from the role they play in supporting evidence-based reasoning.**

This, however, is no easy job. Interpreting the results of data scientific analysis is, more often than not, a highly technical activity and can depart widely from the conventional, everyday styles of reasoning that are familiar to most. Moreover, the various performance metrics deployed in AI systems can be confusing and, at times, seem to be at cross-purposes with each other, depending upon the metrics chosen. There is also an unavoidable dimension of uncertainty that must be accounted for and expressed in confidence intervals and error bars which may only bring further confusion to users and decision subjects.

Be that as it may, by taking a **deliberate and human-centred approach** to the delivery process, you should be able to find the most effective way to convey your model's statistical results to users and decision subjects in non-technical and socially meaningful language that enables them to understand and evaluate the rational justifiability of those results. A good point of departure for this is to divide your map-building task into the *means of content delivery* and the *substance of the content to be delivered*.

*Means of content delivery:* When you start mapping out serviceable ways of presenting and communicating your model's results, you should consider **the users' and decision subjects' perspectives to be of primary importance**. Here are a few guiding questions to ask as you sketch out this dimension of your delivery process as well as some provisional answers to them:

- o How can the delivery process of explaining the AI system's results aid and augment the user's and decision subject's *mental models* (their ways of organising and filtering information), so that they can get a clear picture of the technical meaning of the assessment or explanation? What is the best way to frame the statistical inferences and meanings so that they can be effectively integrated into each user's own cognitive *space of concepts and beliefs*?

While answering these questions will largely depend both on your use case and on the type of AI application you are building, it is just as important that you start responding to them by concentrating on the differing needs and capabilities of your explainees. To do this properly, you should first seek input from domain experts, users, and affected stakeholders, so that you can suitably scan the horizons of existing needs and capabilities. Likewise, you should take a human-centred approach to exploring the types of explanation delivery methods that would best be suited for each of your target groups. Much valuable research has been done on this in the field of human-computer interaction and in the study of human factors. This work should be consulted when mapping delivery means.

Once you have gathered enough background information, you should begin to plan out how you are going to **line up your means of delivery with the varying levels of technical literacy, expertise, and cognitive need possessed by the relevant stakeholder groups, who will be involved in the implementation of your project**. Such a *multi-tiered approach* minimally requires that individual attention be paid to the explanatory needs and capacities of implementers, system operators, and decision subjects and their advocates. This multi-tiered approach will pose different challenges at each different level.

For instance, the mental models of implementers—i.e. their ways of conceptualising the information they are receiving from the algorithmic decision-support system— may, in some cases, largely be shaped by their accumulation of domain know-how and by the filter of on-the-job expertise that they have developed over long periods of practice. These users may have a predisposition to automation distrust or aversion bias, and this should be taken into account when you are formulating appropriate means of explanation delivery.

In other contexts, the opposite may be the case. Where implementers tend to over-rely on or over-comply with automated systems, the means of explanation delivery must anticipate a different sort of mental model and adjust the presentation of information accordingly.

In any event, you will need to have a good empirical understanding of your implementer's decision-making context and maintain such knowledge through ongoing assessment. In both bias risk areas, the conveyance and communication of the assessments generated by algorithmic decision-support systems should attempt to bolster each user's practical judgment in ways that mitigate the possibility of either sort of bias. These assessments should present results as evidence-based reasons that support and better capacitate the objectivity of these implementers' reasoning processes.

The story is different with regard to the cognitive life of the technically inclined user. The mental models of system operators, who are natives in the technical vocabulary and epistemic representations of the statistical results, may be adept at the model-based problem-solving tasks that arise during implementation but less familiar with identifying and responding to the cognitive needs and limitations of non-technical stakeholders. Incorporating ongoing communication exercises and training into their roles in the delivery process may capacitate them to better facilitate implementers' and decision subjects' understanding of the technical details of the assessments generated by algorithmic decision-support systems. These ongoing development activities will not only helpfully enrich operators' mental models, they may also inspire them to develop deeper, more responsive, and more effective ways of communicating the technical yields of the analytics they oversee.

Finally, the mental models of decision subjects and their advocates will show the broadest range of conceptualisation capacities, so your delivery strategy should (1) prioritise the facilitation of optimal explanation at the baseline level of the needs of the most disadvantaged of them and (2) build the depth of your multi-tiered approach to providing effective explanations into the delivery options presented to decision subjects and their advocates. This latter suggestion entails that, beyond provision of the baseline explanation of the algorithmically generated result, options should be given to decision subjects and their advocates to view more detailed and technical presentations of the sort available to implementers and operators (with the proviso that reasonable limitations be placed on transparency in accordance with the need to protect the confidential personal and organisational information and to prevent gaming of the system).

o **How can non-technical stakeholders be adequately prepared to gain baseline knowledge of the kinds of statistical and probabilistic reasoning that have factored into the technical interpretation of the system's output, so that they are able to comprehend it on its own technical terms? How can the technical components be presented in a way that will enable explainees to easily translate the statistical inferences and meanings of the results into understandable and rationally assessable terms? What are the best available media for presenting the technical results in engaging and comprehensible ways?**

To meet these challenges, you should consider supplementing your implementation platform with knowledge-building and enrichment resources that will provide non-technical stakeholders with access to basic technical concepts and vocabulary. At a minimum, you should consider building a plain language glossary of basic terms and concepts that will include all of the technical ideas covered by the algorithmic component of a given explanation. If your explanation platforms are digital, you should also make them as user friendly as possible by hyperlinking the technical terms used in the explanations to their plain language glossary elaborations.

Where possible, explanatory demonstrations of technical concepts (like performance metrics, formal fairness criteria, confidence intervals, etc.) should be provided to users and decision subjects in an engaging and easy-to-comprehend way, and graphical and visualisation techniques should be consistently used to make potentially difficult ideas more accessible. Moreover, the explanation interfaces themselves should be as simple, learnable, and usable as possible. They should be tested to measure the ease with which those with neither technical experience nor domain knowledge are able to gain proficiency in their use and in understanding their content.

**Substance of the technical content to be delivered:** The overall interpretability of your AI system will largely hinge on the effectiveness and even-handedness of your technical content delivery. You will have to strike a balance between (1) determining how best to convey and communicate the rationale of the statistical results so that they may be treated appropriately as decision supporting and clarifying reasons and (2) being clear about the limitations of and potential uncertainties in the statistical results themselves so that the explanations you offer will not mislead implementers and decision subjects. These are not easy tasks and will require substantial forethought as you map out the content clarification aspect of your delivery process.

To assist you in this, here is a non-exhaustive list of recommendations that you should consider as you map out the execution of the technical content delivery component of the responsible implementation of your AI project (This list will, for the sake of specificity, assume the predictive risk assessment example):

- Each explanation should be presented in plain, non-technical language and in an optimally understandable way so that the results provided can enable the affordance of better judgment on the part of implementers and optimal understanding on the part of decision subjects. On the implementer's side, the primary goal of the explanation should be to support the user's ability to offer solid, coherent, and reasonable justifications of their determinations of decision outcomes. On the decision subject's side, the primary goal of the explanation should be to make maximally comprehensible the rationale behind the algorithmic component of the decision process, so that the decision subject can undertake a properly informed critical evaluation of the decision outcome as a whole.

- Each explanation should present its results as facts or evidence in as sparse but complete and sound a manner as possible with a clear indication of what components in the explanation are operating as premises, what components are operating as conclusions, and what the inferential rationale is that is connecting the premises to the conclusions. Each explanation should therefore make explicit the rational criteria for its determination whether this be, for example, global inferences drawn from the population-based reasoning of a demographic analysis or more locally or instance-based inferences drawn from the indication of feature significance by a proxy model. In all cases, the optimisation criteria of the operative algorithmic system should be specified, made explicit, and connected to the logic and rationale of the decision.

- Each explanation should make available the records and activity-monitoring results that the design and development processes of your AI project yielded. Building this link between the process transparency dimension of your project and its outcome transparency will help to make its result, as a whole, more sufficiently interpretable. This can be done by simply linking or including the public-facing component of the process log of your PBG Framework.

- Each explanation provided to an implementer should come with a standard **Implementation Disclaimer** that may read as follows:

<div style="border:1px solid black;">

Implementation Disclaimer:

These results are intended to assist you in making an evidence-based judgment. They are meant neither to replace your reasoned deliberations nor to constitute the sole evidentiary basis of your judgement. These results are also derived from statistical analysis. This means (1) that there are unavoidable possibilities of error and uncertainty in their results, which are specified in the performance measures and confidence intervals provided and (2) that these results are based on population-level data that do not refer specifically to the actual circumstances and abilities of the individual subject of their prediction. The inferences you draw directly from them will therefore be based on statistical generalisation not on an understanding of the life context or concrete potential of the individual person, who will be impacted by your decision.

</div>

- Each explanation should specify and make explicit its governing performance metrics together with the acceptability criteria used to select those metrics and any standard benchmarks followed in establishing that criteria. Where appropriate and possible, fuller information about model validation measurement (including confusion matrix and ROC curve results) and any external validation results should be made available.

- Each explanation should provide confirmatory information that the formal fairness criteria specified in your project's Fairness Policy Statement has been met.

- Each explanation should include clear representations of confidence intervals and error bars. These certainty estimates should make as quantitatively explicit as possible the confidence range of specific predictions, so that users and decision subjects can more fully understand their reliability and the levels of uncertainty surrounding them.

- When an explanation offers categorically ordered scores (for instance, risk scores on a scale of 1 to 10), that explanation must also explicitly indicate the actual raw numerical probabilities for the labels (predicted outcomes) that have been placed into those categories. This will help your delivery process avoid producing confusion about the relative magnitudes of the categorical groupings under which the various scores fall. Information should also be provided about the relative distances between the risk scores of specific cases if the risk categories under which they are placed are unevenly distributed. It may be possible, for example, for two cases, which fall under the same high risk category (say, 9) to be farther apart in terms of the actual values of their risk probabilities than two other cases in two different categories (say 1 and 4). This may be misleading to the user.

- Each explanation should, where possible, include a counterfactual explanatory tool, so that implementers and affected individuals have the opportunity to gain a better contrastive understanding of the logic of the outcome and its alternative possibilities.

**Social component of responsible implementation:** We have now established the first step in the delivery of a responsible implementation process: making clear the rationale behind the technical content of an algorithmic model's statistical results and determining how best to convey and communicate it so that these results may be appropriately treated as decision supporting and clarifying reasons. This leaves us with a second related task of content clarification, which is only implicit in the first step but must be made explicit and treated reflectively in a second.

Beyond translating statistically expressed results into humanly significant reasons, you will have to make sure that their *socially meaningful content* is clarified by implementers, so that they are able to thoughtfully apply these results to the real human lives they impact in terms of the specific societal and individual contexts in which those lives are situated.

This will involve explicitly translating that model's technical machinery—its input and output variables, parameters, and functional rationale—*back* into the everyday language of the humanly relevant meanings, categories, and relationships that informed the formulation of its purpose, objectives, and intended elements of design in the first place. It will also involve training and preparing implementers to intentionally assist in carrying out this translation in each particular case, so that due regard for the dignity of decision subjects can be supported by the interpretive charity, reasonableness, empathy, and context-specificity of the determination of the outcomes that affect them.

Only through this re-translation will the internals, mechanisms, and output of the model become *useably interpretable* by implementers: Only then will they be able to apply input features of relevance to the specific situations and attributes of decision subjects. Only then will they be able to critically assess the manner of inference-making that led to its conclusion. And only then will they be able to adequately weigh the normative considerations (such as prioritising public interest or safeguarding individual well-being) that factored into the system's original objectives.

Having clarified the socially meaningful content of the model's results, the implementer will be able to more readily apply its evidentiary contribution to a more holistic and wide-angled consideration of the particular circumstances of the decision subject while, at the same time, weighing these circumstances against the greater purpose of the algorithmically assisted assessment. It is important to note here that the understanding enabled by the clarification of the social context and stakes of an algorithmically supported decision-making process goes hand-in-glove with fuller considerations of the moral justifiability of the outcome of that process.

A good starting point for considering how to integrate this clarification of the socially meaningful content of an algorithmic model's output into your map of the delivery process is to consider what you might think of as your AI project's **content lifecycle**.

**The content lifecycle**: The output of an algorithmic system does not begin and end with the computation. Rather, it begins with the very human purposes, ideas, and initiatives that lay behind the conceptualisation and design of that system. Creating technology is a shared public activity, and it is animated by human objectives and beliefs. An algorithmic system is brought into the world as the result of this collective enterprise of ingenuity, intention, action, and collaboration.

Human choices and values therefore punctuate the design and implementation of AI systems. These choices and values are inscribed in algorithmic models:

- At the very inception of an AI project, human choices and values come into play when we formulate the goals and objectives to be achieved by our algorithmic technologies. They come into play when we define the optimal outcome of our use of such technologies and when we translate these goals and objectives into target variables and their measurable proxies.

- Human choices and values come into play when decisions are made about the sufficiency, fit-for-purpose, representativeness, relevance, and appropriateness of the data sampled. They come into play in how we curate our data—in how we label, organise, and annotate them.

- Such choices and values operate as well when we make decisions about how we craft a feature space—how we select or omit and aggregate or segregate attributes. Determinations of what is relevant, reasonable, desirable, or undesirable will factor into what kinds of inputs we are going to include in the processing and how we are going to group and separate them.

- Moreover, the data points themselves are imbued with residua of human choices and values. They carry forward historical patterns of social and cultural activity that may contain configurations of discrimination, inequality, and marginalisation—configurations that must be thoughtfully and reflectively considered by implementers as they incorporate the analytics into their reasoned determinations.

Whereas all of these human choices and values are translated in to the algorithmic systems we build, the responsible implementation of these systems requires that they be translated out. The rationale and logic behind an algorithmic model's output can be properly understood as it affects the real existence of a decision subject only when we transform its variables, parameters, and analytical structures back into the human currency of values, choices, and norms that shaped the construction of its purpose, its intended design, and its optimisation logic from the start.

It is only in virtue of this **re-translation** that an algorithmically supported outcome can afford stakeholders the degree of deliberation, dialogue, assessment, and mutual understanding that is necessary to make it fully comprehensible and justifiable to them. And, it is likewise only in virtue of this re-translation that the implementation process itself can, at once, secure end-to-end accountability and give due regard to the SUM values.

The content lifecycle of algorithmic systems therefore has three phases: (1) The **translation in** of human purposes, values, and choices during the design process; (2) The digital processing of the quantified/mechanised proxies of these purposes, values, and choices in the statistical frame; (3) The **translation out** of the purposes, values, and choices in clarifying the socially meaningful content of the result as it affects the life of the decision subject through the implementation process. Here is a visualisation of these three phases of the content lifecycle:

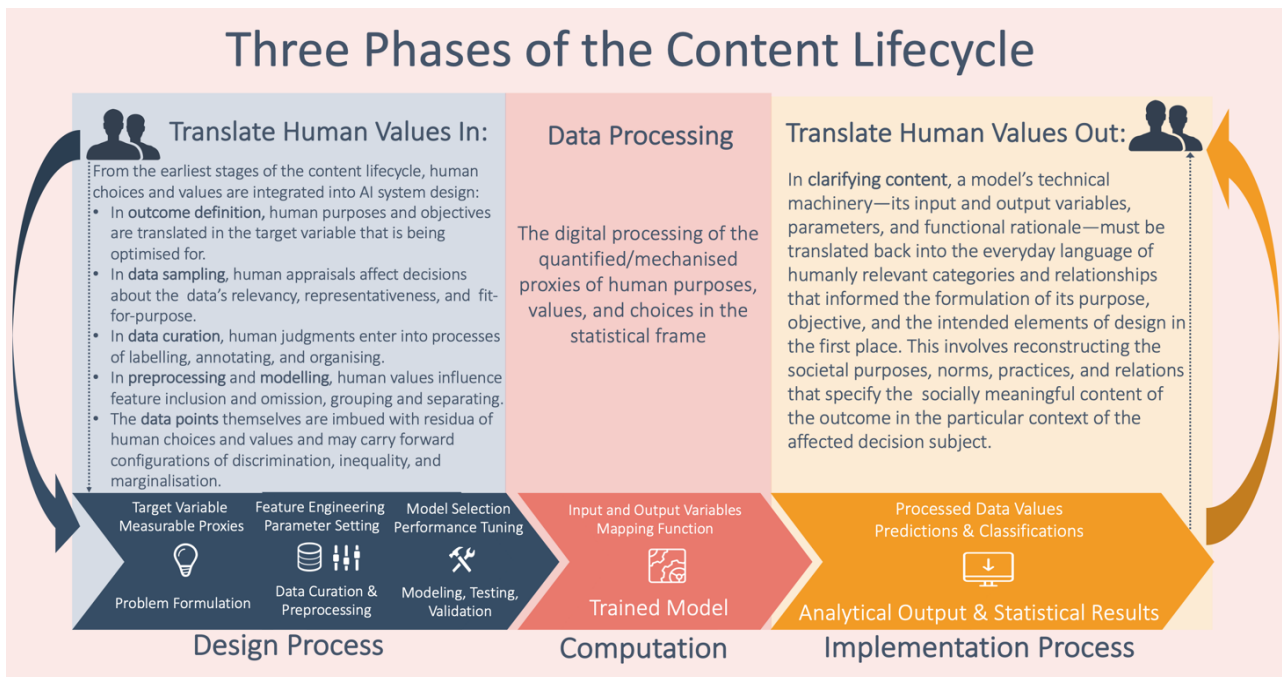

**The translation rule:** A beneficial result of framing the implementation process in terms of the content lifecycle is that it gives us a clear and context-sensitive measure by which to identify the explanatory needs of any given AI application. We can think of this measurement as the **translation rule**. It states that:

**What is *translated in* to an algorithmic system with regard to the human choices and societal values that determine its content and purpose is directly proportional to what, in terms of the explanatory needs of clarification and justification, must be *translated out*.**

The translation rule organically makes two distinctions that have great bearing on the delivery process for responsible implementation. First, it divides the question of what needs explaining into two parts: (1) issues of socially meaningful content in need of clarification (i.e., the explanatory need that comes from the **translation in** to the AI model of the categories, meanings, and relations that originate in social practices, beliefs, and intentions) (2) issues of normative rightness in need of justification (i.e. the explanatory need that comes from **translation in** to the AI model of choices and considerations that have bearing on its ethical permissibility, discriminatory non-harm, and public trustworthiness). These two parts line up with what we have <u>above</u> called **interpretable AI** and **justifiable AI** respectively, and what we have also identified as <u>tasks 2 and 3</u> of delivering transparent AI.

Secondly, the translation rule divides the two dimensions of translation (translation in and translation out) into aspects of **intention-in-design** and **intention-in-application**. *Translating in* has to do with **intention-in-design**. It involves an active awareness of the human purposes, objectives, and intentions that factor into the construction of AI systems. **Translating out,** on the other hand, has directly to do with **intention-in-application**, or put differently, the intentional dimension of the implementation of an AI system by a user in a specific context and with direct consequences for a subject affected by its outcome.

In human beings, intention-in-design and intention-in-application are **united in intelligent action**, and it is precisely this unity that enables people to reciprocally hold each other accountable for the consequences of what they say and what they do. By contrast, in artificial intelligence systems, which fulfil surrogate cognitive functions in society but are themselves neither intentional nor accountable, design and application are divided. In these systems, intention-in-design and intention-in-application are and must remain **punctuation points of human involvement and responsibility** that manifest on either side of the vacant mechanisms of data processing. This is why translation is so important, and this is why enabling the implementer's capacity to **intentionally translate out the social and normative content** of the model's results is such a critical element of the responsible delivery of your AI project.

It might be helpful to think more concretely about the translation rule by considering it in action. Let's compare two hypothetical examples: (1) a use case about an early cancer detection system in radiomics (a machine learning application that uses high throughput computing to identify features of pathology that are undetectable to the trained radiological eye); and (2) a use case about a predictive risk assessment application that supports decision-making in child social care.

In the radiomics case, the **translating in** dimension involves minimal social content: the clinical goal inscribed in the model's objective is that of lesion detection and the features of relevance are largely voxels extracted from PET and CT scanner images. However, the normative aspect of **translating in** is, in this case, significant. Ethical considerations about looking after patient wellbeing and clinical safety are paramount and wider justice concerns about improving healthcare for all and health equity factor in as well.

The explanatory needs of the physician/implementer receiving clinical decision support and of the clinical decision subject will thus lean less heavily on the dimension of the clarification of socially meaningful content than it will on the normative dimension of justifying the safety of the system, the priority of the patient's wellbeing, and the issues of improved delivery and equitable access. The technical content of the decision support may be crucial here (Issues surrounding the reproducibility of the results and the robustness of the system may, in fact, be of great concern in the assessment of the validity of the outcome.), but the **translating out** component of the implementation remains directly proportional to the minimal social content and to the substantial ethical concerns and objectives that were **translated in** and that thus inform the explanatory and justificatory needs of the result in general.

The explanatory demands in the child social care risk assessment use case are entirely different. The social content of the **translating in** dimension is intricate, multi-layered, and extensive. The chosen target variable may be child safety or the prevention of severe mistreatment and the measurable proxy, home removal of at-risk children within a certain timeframe. Selected features that are deemed relevant may include the age of the at-risk children, public health records, previous referrals, family history of violent crime, welfare records, juvenile criminal records, demographic information, and mental health records. Complex socioeconomic and cultural formations may additionally influence the representativeness and quality of the dataset as well as the substance of the data itself.

The normative aspect of *translating in* here is also subtle and complicated. Ethical considerations about protecting the welfare of children at risk are combined with concerns that parents and guardians be treated fairly and without discrimination. Objectives of providing evidence-based decision support are also driven by hopes that accurate results and well-reasoned determinations will preserve the integrity and sanctity of familial relations where just, safe, and appropriate. Other goals and purposes may be at play as well such as making an overburdened system of service provision more efficient or accelerating real-time decision-making without harming the quality of the decisions themselves.

In this case of predictive risk assessment, the *translating out* burdens of the frontline social worker are immense both in terms of clarifying content and in terms of moral justification. If, for example, analytical results yielding a high risk score were based on the relative feature importance of demographic information, welfare records, mental health records, and criminal history, the implementer would have to scrutinise the particular decision subject's situation, so that the socially meaningful content of these factors could be clarified in terms of the living context, relevant relationships, and behavioural patterns of the stakeholders directly affected. Only then could the features of relevance be thoroughly and deliberatively assessed.

The effective interpretability of the model's result would, in this case, heavily depend on the implementer's ability to apply domain-knowledge in order to reconstruct the meaningful social formations, intentions, and relationships that constituted the concrete form of life in which the predictive risk modelling applies. The implementer's well-reasoned decision here would involve a careful weighing of this socially clarified content against the wider predictive patterns in the data distribution yielded by the model's results—patterns that may have otherwise gone unnoticed.

Such a weighing process would, in turn, be informed by the normative-explanatory need to **translate out** the morally implicating choices, concerns, and objectives that influenced and informed the predictive risk assessment model's development in the first place. Again, the interpretive burden of the frontline social worker would be immense here. First, this implementer would have to deliberate with a critically informed awareness of the legacies of discrimination and inequity that tend to feed forward in the kinds of evidentiary sources drawn upon by the analytics. Such an active reflexivity is crucial for retaining the punctuating role of human involvement and responsibility in these sensitive and high-stakes environments.

Just as importantly, the frontline social worker would have to evaluate the real impact of ethical objectives at the point of delivery. Not only would the results of the analytics have to be aligned with the ethical concerns and purposes that fostered the construction of the model, this implementer would have to reflectively align their own potentially diverging ethical point of view both with those results and with those objectives. This ***normative*** *triangulation* between the original intention-in-design, the implementer's intention-in-application, and the content clarification of the AI system's results is, in fact, a crucial safeguard to the delivery of justifiable AI. It again enables a reanimation of moral involvement and responsibility at the most critical juncture of the content lifecycle.

Step 3: Build an ethical implementation platform:

☐ (1) **Train ethical implementation**. The continuous challenges of translation, content clarification, and normative explanation should inform how you set up your implementation training to achieve optimal outcome transparency. In addition to the necessary <u>training to prevent implementation biases in the users of your AI system</u> (discussed above), you should prepare and train the implementers to be stewards of interpretable and justifiable AI. This entails that they be able to:

- o Rationally evaluate and critically assess the logic and rationale behind the outputs of the AI systems;

- o Convey and communicate their algorithmically assisted decisions to the individuals affected by them in plain language. This includes explaining to them in an everyday, non-technical, and accessible way how and why the decision-supporting model performed the way it did in a specific context and how that result factored into the final outcome of the implementation;

- o Apply the conclusions reached by the AI model to a more focused consideration of the particular social circumstances and life context of the decision subject and other affected parties;

- o Treat the inferences drawn from the results of the model's computation as evidentiary contributions to a broader, more rounded, and coherent understanding of the individual situations of the decision subject and other affected parties;

- o Weigh the interpretive understanding gained by integrating the model's insights into this rounded picture of the life context of the decision subject against the greater purpose and societal objective of the algorithmically assisted assessment;

- o Justify the ethical permissibility, the discriminatory non-harm, and the public trustworthiness both of the AI system's outcome and of the processes behind its design and use

☐ (2) **Make your implementation platform a relevant part and capstone of the sustainability track of your project**. An important element of gauging the impacts of your AI technology on the individuals and communities it touches is having access to the frontlines of its potentially transformative and long-term effects. Your implementation platform should assist you in gaining this access by being a *two-way medium of application and communication*. It should both enable you to sustainably achieve the objectives and goals you set for your project through responsible implementation, but it should also be a sounding board as well as a site for feedback and cooperative sense-checking about the real-life effects of your system's use.

Your implementation platform should be dialogically and collaboratively connected to the stakeholders it effects. It should be bound to the communities it serves as part of a shared project to advance their immediate and long-run wellbeing.

☐ **(3) Provide a model sheet to implementers and establish protocols for implementation reporting.** As part of the roll-out of your AI project, you should prepare a summary/model sheet for implementers, which includes summation information about the system's technical specifications and all of the relevant details indicated above in the section on *substance of the technical content to be delivered*. This should include relevant information about performance metrics, formal fairness criteria and validation, the implementation disclaimer, links or summaries to the relevant information from the process logs of your PBG Framework, and links or summary information from the Stakeholder Impact Assessment.

You should also set up protocols for implementation reporting that are proportional to the potential impacts and risks of the system's use.

☐ **(4) Foster outcome understanding through dialogue.** Perhaps the single most important aspect of building a platform for ethical implementation is the awareness that the realisation of interpretable and justifiable AI is a dialogical and collaborative effort. Because all types of explanation are mediated by language, each and every explanatory effort is a participatory enterprise where understanding can be reached only through acts of communication. The interpretability and justifiability of AI systems depend on this shared human capacity to give and ask for reasons in the ends of reaching mutual understanding. Implementers and decision subjects are, in this respect, first and foremost participants in an explanatory dialogue, and the success of their exchange will hinge both on a reciprocal readiness to take the other's perspective and on a willingness to enlarge their respective mental models in accordance with new, communicatively achieved, insights and understandings.

For these reasons, your implementation platform should encourage open, mutually respectful, sincere, and well-informed dialogue. Reasons from all affected voices must be heard and considered as demands for explanation arise, and manners of response and expression should remain clear, straightforward, and optimally accessible. Deliberations that have been inclusive, unfettered, and impartial tend to generate new ideas and insights as well as better and more inferentially sound conclusions, so approaching the interpretability and justifiability of your AI project in this manner will not only advance its responsible implementation, it will likely encourage further improvements in its design, delivery, and performance.

# Conclusion:

In 1936, a 23-year-old mathematician from Maida Vale named Alan Turing sat down with pencil and paper. Using just the image of a linear tape divided evenly into squares, a list of symbols, and a few basic rules, he drew a sketch to show the step-by-step process of how a human being can carry out any calculation, from the simplest operation of arithmetic to the most complex nonlinear differential equation.

Turing's remarkable invention (now known simply as the Turing machine) solved the perplexing and age-old mathematical question of *what an effective calculation is*—the question of *how to define an algorithm*. Not only did Turing show what it means to compute a number by showing *how humans do it*, he created, in the process, the idea behind the modern general purpose computer. Turing's astonishingly humble innovation ushered in the digital age.

Just over eight decades later, as we step forward together into the open horizons of a rapidly evolving digital future, it is difficult to image that what started as a thought experiment in a small room at Kings College, Cambridge has now become such a humanly defining force. We live in an increasingly dynamic and integrated computational reality where connected devices containing countless sensors and actuators intermingle with omnipresent algorithmic systems and cloud computing platforms.

With the rise of the Internet of Things, edge computing, and the expanding smart automation of infrastructure, industry, and the workplace, AI systems are progressively more coming to comprise the cyber-physical frame and fabric of our networked society. For better or worse, artificial intelligence is not simply becoming a general purpose technology (like steam power or electricity). It is, more essentially, becoming a gatekeeper technology that uniquely holds the key both to the potential for the exponential advancement of human wellbeing and to possibilities for the emergence of significant risks for society's future. It is, as yet, humankind that must ultimately choose which direction the key will turn.

This choice leaves difficult questions in the lap of the moral agency of the present: What shape will the data-driven society of tomorrow take? How will the values and motivations that are currently driving the gathering energies of technological advancement in artificial intelligence come both to influence that future society's ways of life and to transform the identities of its warm-blooded subjects?

This guide on understanding AI ethics and safety has offered you one way to move forward in answering these questions. In a significant sense, it has attempted to prepare you to take Turing's lead: to see the design and implementation of algorithmic models as an eminently *human activity*— an activity guided by our purposes and values, an activity for which, each of us, who is involved in the development and deployment of AI systems, is morally and socially responsible.

This starting point in human action and intention is a crucial underpinning of responsible innovation. For, it is only when we prioritise considerations of the ethical purposes and values behind the trajectories of our technological advancement, that we, as vested societal stakeholders, will be able to take the reins of innovation and to steer the course of our algorithmic creations in accordance with a shared vision of what a better human future should look like.

# Acknowledgments:

Writing this guide would simply not have been possible without the hard work, dedication, and insight of so many interlocutors both within The Alan Turing Institute and through the meaningful partnerships that the Turing's Public Policy Programme has formed with stakeholders from across the UK Government.

To take the latter group first, the Office for Artificial Intelligence (OAI) and the Government Digital Service (GDS)'s keen vision and their commitment to responsible AI innovation have been an enabling condition of the development of this work. In particular, the patience and incisiveness of OAI's Sebastien Krier and Jacob Beswick, and GDS' Bethan Charnley have been instrumental in bringing the project to its completion.

I am also incredibly grateful for the impact that our interactions with the Ministry of Justice (MoJ)'s Data Science Hub has had on developing the framing for this guide. Input from the MoJ's Megan Whewell, Philip Howard, Jonathan Roberts, Olivia Lewis, Ross Wyatt, and from its Data Science Innovation Board have left a significant mark on the research.

Last, but not least, our ongoing partnership with the Information Commissioner's Office on Project Expl*AI*n—and, in particular, with ICO colleagues Carl Wiper and Alex Hubbard—has been a key contributor to this guide's focus on fairness, transparency, and accountability. Project Expl*AI*n aims to provide practical guidance for organisations on explaining AI supported decisions to the subjects of those decisions. Taking inspiration from our work on Project Expl*AI*n and from the input gathered over the course of the two citizens' juries we held in Manchester and Coventry, the current guide emphasises the importance of communication and attempts to build out a vision of human-centred and context-sensitive implementation.

As the Ethics Fellow within the Public Policy Programme at the Turing, I have benefited tremendously from being surrounded by an immensely talented group of thinkers and doers, whose commitment to making the connected world a better place through interdisciplinary research and advisory intervention is an inspiration every day. Programme Director, Helen Margetts, and Deputy Director, Cosmina Dorobantu, have been crucial and inimitable supports of this project from its inception as has my small but brilliant team of researchers, Josh Cowls and Christina Hitrova. My involvement with the Turing's Data Ethics Group has also been a tremendous source of insight and inspiration for this project. Given the ambitious deadlines that accompanied this guide's final stages of production, heroic efforts to review its contents as a whole or in parts were made by Florian Ostmann, Michael Veale, David Watson, Mark Briers, Evelina Gabsova, Alexander Harris, and Anna FitzMaurice. Their perceptive feedback notwithstanding, any unclarities that appear in *Understanding Artificial Intelligence Ethics and Safety* reflect the faults of its author alone.

# Bibliography and Further Readings

Included here is a bibliography organised into the main themes covered in this guide. Please use this as a starting point for further exploration of these complex topics. Many thanks to the tireless efforts of Jess Morley and Corianna Moffatt without whom this bibliography could not have been compiled.

## The SUM Values

## General fairness

## Data fairness

Varshney, K. R., & Alemzadeh, H. (2017). On the safety of machine learning: Cyber-physical systems, decision sciences, and data products. *Big Data*, *5*(3), 246-255. Retrieved from https://www.liebertpub.com/doi/abs/10.1089/big.2016.0051

Vidgen, B., Nguyen, D., Tromble, R., Hale, S., Margetts, H., Harris, A. (2019) 'Challenges and frontiers in abusive content detection', *Forthcoming* ACL 2019.

Zheng, X., Wang, M., & Ordieres-Meré, J. (2018). Comparison of data preprocessing approaches for applying deep learning to human activity recognition in the context of industry 4.0. *Sensors*, *18*(7), 2146. Retrieved from https://www.mdpi.com/1424-8220/18/7/2146

## Design fairness

Barocas, S., & Selbst, A. D. (2016). Big Data's disparate impact. *Calif. L. Rev.*, *104*, 671. Retrieved from https://heinonline.org/hol-cgi-bin/get_pdf.cgi?handle=hein.journals/calr104§ion=25

Calders, T., & Verwer, S. (2010). Three naive Bayes approaches for discrimination-free classification. *Data Mining and Knowledge Discovery*, *21*(2), 277–292. https://doi.org/10.1007/s10618-010-0190-x

Calmon, F., Wei, D., Vinzamuri, B., Ramamurthy, K. N., & Varshney, K. R. (2017). Optimized pre-processing for discrimination prevention. In *Advances in Neural Information Processing Systems* (pp. 3992-4001). Retrieved from http://papers.nips.cc/paper/6988-optimized-pre-processing-for-discrimination-prevention

d'Alessandro, B., O'Neil, C., & LaGatta, T. (2017). Conscientious classification: A data scientist's guide to discrimination-aware classification. *Big Data*, *5*(2), 120-134. https://doi.org/10.1089/big.2016.0048

Hajian, S., Bonchi, F., & Castillo, C. (2016). Algorithmic bias: From discrimination discovery to fairness-aware data mining. In *Proceedings of the 22nd ACM SIGKDD international conference on knowledge discovery and data mining* (pp. 2125-2126). ACM. Retrieved from https://dl.acm.org/citation.cfm?id=2945386

Kamiran, F., & Calders, T. (2012). Data preprocessing techniques for classification without discrimination. *Knowledge and Information Systems*, *33*(1), 1-33. Retrieved from https://link.springer.com/article/10.1007/s10115-011-0463-8

Lehr, D., & Ohm, P. (2017). Playing with the data: What legal scholars should learn about machine learning. *UCDL Rev.*, *51*, 653. Retrieved from https://lawreview.law.ucdavis.edu/issues/51/2/Symposium/51-2_Lehr_Ohm.pdf

Passi, S., & Barocas, S. (2019). Problem formulation and fairness. In *Proceedings of the Conference on Fairness, Accountability, and Transparency* (pp. 39-48). ACM. Retrieved from https://dl.acm.org/citation.cfm?id=3287567

Singh, J., & Sane, S. S. (2014). Preprocessing technique for discrimination prevention in data mining. *The IJES, 3*(6), 12-16. Retrieved from https://www.academia.edu/6994180/Pre-Processing_Approach_for_Discrimination_Prevention_in_Data_Mining

Singhal, S., & Jena, M. (2013). A study on WEKA tool for data preprocessing, classification and clustering. *International Journal of Innovative technology and exploring engineering (IJItee)*, *2*(6), 250-253. Retrieved from https://pdfs.semanticscholar.org/095c/fd6f1a9dc6eaac7cc3100a16cca9750ff9d8.pdf

van der Aalst, W. M., Bichler, M., & Heinzl, A. (2017). Responsible data science. *Springer Fachmedien Wiesbaden.* https://doi.org/10.1007/s12599-017-0487-z

## Outcome fairness

Agarwal, A., Beygelzimer, A., Dudík, M., Langford, J., & Wallach, H. (2018). A reductions approach to fair classification. *ArXiv:1803.02453*. Retrieved from http://arxiv.org/abs/1803.02453

## Implementation fairness

## Accountability

## Stakeholder Impact Assessment

**Hong Kong**

The Information Accountability Foundation. (2018a). *Ethical accountability framework for Hong Kong, China: A report prepared for the Office of the Privacy Commission for Personal Data.* Retrieved from https://www.pcpd.org.hk/misc/files/Ethical_Accountability_Framework.pdf

The Information Accountability Foundation. (2018b). *Data stewardship accountability, data impact assessments and oversight models: Detailed support for an ethical accountability framework.* Retrieved from https://www.pcpd.org.hk/misc/files/Ethical_Accountability_Framework_Detailed_Support.pdf

**Canada**

Treasury Board of Canada Secretariat. (2019). *Algorithmic impact assessment.* Retrieved from https://open.canada.ca/data/en/dataset/748a97fb-6714-41ef-9fb8-637a0b8e0da1

**Safety: Accuracy, reliability, security, and robustness**

Amodei, D., Olah, C., Steinhardt, J., Christiano, P., Schulman, J., & Mané, D. (2016). Concrete problems in AI safety. *arXiv:1606.06565*. Retrieved from https://arxiv.org/abs/1606.06565

Auernhammer, K., Kolagari, R. T., & Zoppelt, M. (2019). Attacks on Machine Learning: Lurking Danger for Accountability [PowerPoint Slides]. Retrieved from https://safeai.webs.upv.es/wp-content/uploads/2019/02/3.SafeAI.pdf

Demšar, J., & Bosnić, Z. (2018). Detecting concept drift in data streams using model explanation. *Expert Systems with Applications*, *92*, 546–559. https://doi.org/10.1016/j.eswa.2017.10.003

Google. (2019). *Perspectives on issues in AI governance.* Retrieved from https://ai.google/static/documents/perspectives-on-issues-in-ai-governance.pdf

Göpfert, J. P., Hammer, B., & Wersing, H. (2018). Mitigating concept drift via rejection. In *International Conference on Artificial Neural Networks* (pp. 456-467). Springer, Cham. https://doi.org/10.1007/978-3-030-01418-6_45

Irving, G., & Askell, A. (2019). AI safety needs social scientists. *Distill*, *4*(2). https://doi.org/10.23915/distill.00014

Kohli, P., Dvijotham, K., Uesato, J., & Gowal, S. (2019). Towards a robust and verified AI: Specification testing, robust training, and formal verification. *DeepMind Blog.* Retrieved from https://deepmind.com/blog/robust-and-verified-ai/

Kolter, Z., & Madry, A. (n.d.). Materials for tutorial adversarial robustness: Theory and practice. Retrieved from https://adversarial-ml-tutorial.org/

Marcus, G. (2018). Deep learning: A critical appraisal. *arXiv:1801.00631*. Retrieved from https://arxiv.org/abs/1801.00631

Muñoz-González, L., Biggio, B., Demontis, A., Paudice, A., Wongrassamee, V., Lupu, E. C., & Roli, F. (2017, November). Towards poisoning of deep learning algorithms with back-gradient optimization. In *Proceedings of the 10th ACM Workshop on Artificial Intelligence and Security* (pp. 27-38). ACM. Retrieved from https://dl.acm.org/citation.cfm?id=3140451

Nicolae, M. I., Sinn, M., Tran, M. N., Rawat, A., Wistuba, M., Zantedeschi, V., ... & Edwards, B. (2018). Adversarial Robustness Toolbox v0.4.0. *arXiv:1807.01069*. Retrieved from https://arxiv.org/abs/1807.01069

Ortega, P. A., & Maini, V. (2018). Building safe artificial intelligence: specification, robustness, and assurance. *DeepMind Safety Research Blog, Medium*. Retrieved from https://medium.com/@deepmindsafetyresearch/building-safe-artificial-intelligence-52f5f75058f1

## Transparency

## Process-Based Governance

## Interpretable AI

Responsible delivery through human-centred implementation protocols and practices

## Individual and societal impacts of machine learning and algorithmic systems

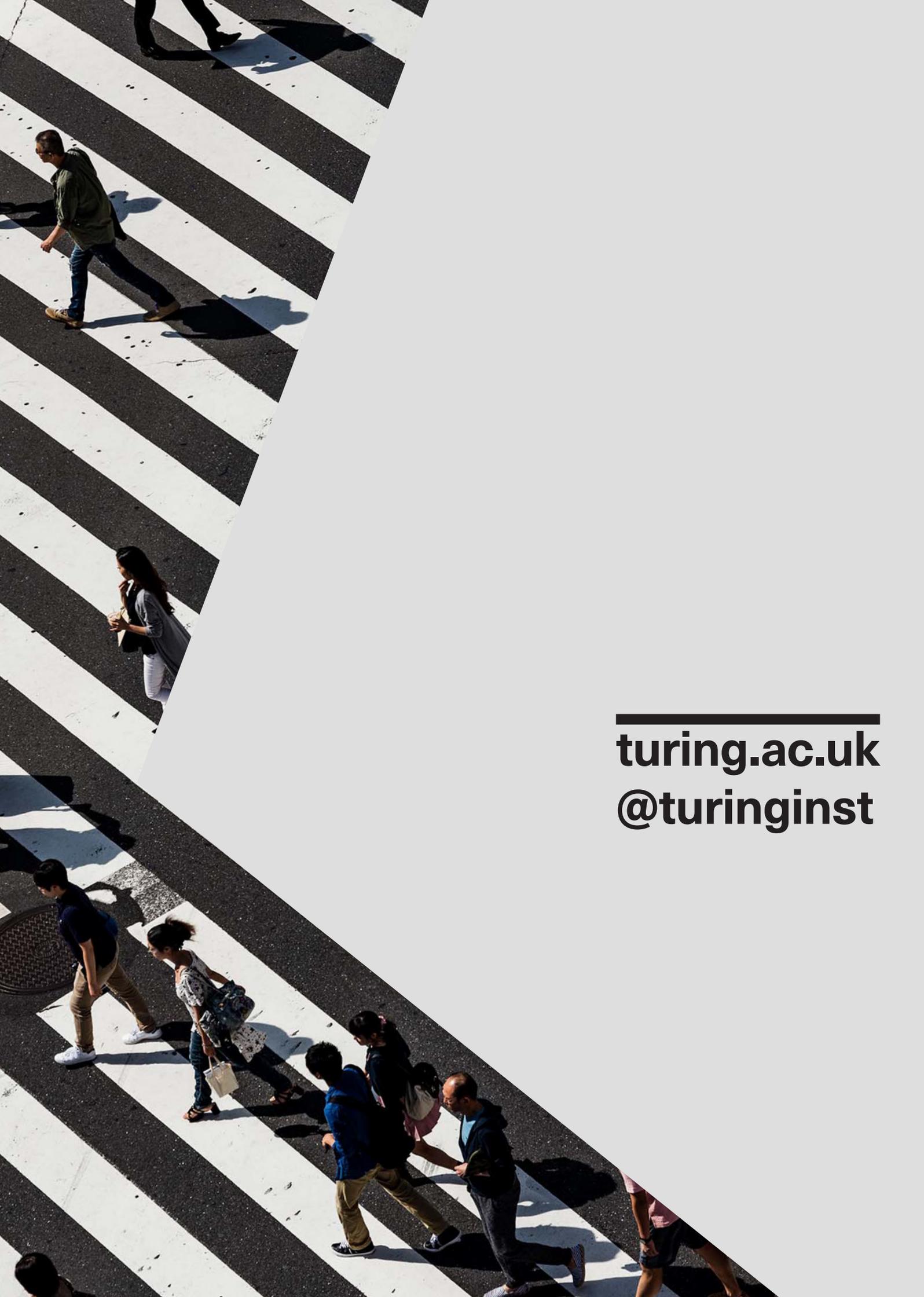

turing.ac.uk
@turinginst